\documentclass[preprint]{aastex}
\slugcomment{submitted to {\it The Astronomical Journal}}
\begin{document}
\title{WIYN Open Cluster Study. XXXIX.  Abundances in NGC 6253 from HYDRA Spectroscopy of the Li 6708 \AA\ Region}  
%\title{HYDRA Spectroscopy of NGC 6253: I. Abundances from the Li 6708 \AA\ Region}
\author{Barbara J. Anthony-Twarog$^{1,3}$, Constantine P. Deliyannis$^{2,3}$, Bruce A. Twarog$^{1,3}$, Jeffrey D. Cummings$^{2,3}$, \& Ryan M. Maderak$^2$}
\affil{$^1$Department of Physics and Astronomy, University of Kansas, Lawrence, KS 66045-7582}
\affil{$^2$Department of Astronomy, Indiana University, Bloomington, IN 47405-7105}
\affil{Electronic mail: bjat@ku.edu, con@astro.indiana.edu, btwarog@ku.edu, jdcummi@astro.indiana.edu, maderak@astro.indiana.edu}
\altaffiltext{3}{Visiting Astronomer, Cerro Tololo Inter-American Observatory. CTIO is operated by AURA, Inc., under contract to the National 
Science Foundation.}

\begin{abstract}
High-dispersion spectra of 89 potential members of the old, super-metal-rich open cluster, NGC 6253, have been obtained with the 
HYDRA multi-object spectrograph. Based upon radial-velocity measurements alone, 47 stars at the turnoff of the cluster color-magnitude 
diagram (CMD) and 18 giants are identified as potential members. Five turnoff stars exhibit evidence of binarity while proper-motion 
data eliminates two of the dwarfs as members. The mean cluster radial velocity from probable single-star members 
is -29.4 $\pm$ 1.3 km/sec (sd). A discussion of the current estimates for the cluster reddening, derived independently of potential 
issues with the $BV$ cluster photometry, lead to an adopted reddening of E$(B-V)$ = 0.22 $\pm$ 0.04. From equivalent width 
analyses of 38 probable single-star members near the CMD turnoff, the weighted average abundances are found 
to be [Fe/H] = $+0.43 \pm 0.01$, [Ni/H] = $+0.53 \pm 0.02$ and [Si/H] = $+0.43 ^{0.03}_{0.04}$, where the errors refer to 
the standard errors of the weighted mean. Weak evidence is found for a possible decline in metallicity with increasing luminosity
among stars at the turnoff. We discuss the possibility that our turnoff stars have been affected by microscopic diffusion.
For 15 probable single-star members among the giants, spectrum synthesis leads to abundances of 
$+0.46^{0.02}_{0.03}$ for [Fe/H]. While less than half the age of NGC 6791, NGC 6253 is at least as metal-rich and, within the uncertainties, 
exhibits the same general abundance pattern as that typified by super-metal-rich dwarfs of the galactic bulge. 
\end{abstract}

\keywords{open clusters and associations: individual (NGC 6253, NGC 6791) - stars: abundances - techniques: spectroscopic}

\section{INTRODUCTION}
Given the amount and level of detailed information potentially available for the local region of the Galaxy, the chemical and dynamical evolution 
of the Galactic disk offers unrivalled insight into the process of galaxy formation and evolution, at least in the context of generating a 
well-defined spiral galaxy as the end product. As the database for the population components comprising the Galaxy within the neighborhood
of the Sun and beyond has grown in size and precision, the simple picture of \citet{els} has given way to a complex evolutionary history
increasingly dominated by multiple substructures from within and multiple mergers from without (see, e.g., \citet{pra08, fre08, wys08} among many
others). To make sense of these complications, it is valuable to view the Sun and the solar neighborhood within the context of the temporal and
spatial evolution of the entire disk. The delineation of that global context requires age, metallicity, and kinematic measures for systems 
that inhabit zones of the Galaxy well within and well beyond the solar circle. While they can be challenging observationally and their number 
is modest, open clusters increasingly have been selected as the objects of choice in attempting to probe the history of the Galaxy at large 
distance and over long periods of time \citep{bra08, fri08}. The primary driver behind this trend is the often-noted benefit of being able to 
derive the age, composition, and distance from collective analysis of the homogeneous cluster stellar sample, rather than from an individual star. 
The challenges in interpreting the results from the cluster population are often coupled to small samples over a wide range in parameter 
space \citep{car98}, a lack of homogeneity in merging the data from different observers applying different approaches \citep{che03}, and 
the real possibility that subgroups of clusters reflect different origins indicative of the complex structure noted above for field 
stars \citep{fri06, car06, war09}. 

Among the various properties of the disk that can be probed using open clusters, two primary questions of interest have been the existence of
an age-metallicity relation (AMR) \citep{pia95, car98} and the delineation of a radial metallicity gradient \citep{fri95, taat, car98, fri02, che03}. 
Once one removes the effect of the latter from the sample, the conclusions regarding the absence of the former have changed little over the 
last 30 years \citep{hir78, mcc78, fri95, fri02}. However, the failure to discern an AMR among open clusters may be partly due to the limited 
age range contained within most cluster samples. Numerous studies over the last few decades have concluded that there is little evidence 
for a significant change in the mean [Fe/H] among stars formed within the solar neighborhood over the last 5-6 Gyrs 
\citep{twa80, meu91, edv, gar00, fel01, nor04}, though exceptions to this pattern occasionally arise \citep{roc00, sou08}. Since few clusters 
survive to this age due to tidal disruption, most cluster-based determinations of the AMR are dominated by samples where no trend is expected, 
assuming clusters reflect the same history as the field stars. By contrast, if the absence of an AMR over the last 5-6 Gyrs holds throughout 
the disk, the abundance gradient of the open cluster sample should be approximately constant with time and, within the statistical uncertainty, 
should not be dependent upon the mix of ages included within the cluster analysis. As cluster studies extend the temporal and spatial 
baseline of the sample, the relevance of the objects that populate the extreme ends of the scales becomes an issue of greater concern. For 
example, the observation that the open cluster sample beyond a galactocentric distance of 10 kpc, on a scale where the Sun is positioned at 8.5 kpc, 
exhibits, at best, a shallow gradient in metallicity \citep{taat} has now been confirmed through the addition of clusters beyond 20 kpc from the galactic 
center \citep{yon05, ses08}. This immediately raises the question of whether or not the very distant members of this sample are, in fact, 
representative of the internal development and evolution of the outer Galactic disk or are interlopers accreted via interactions with nearby galactic 
companions. It should be apparent, however, that this dichotomy may be artificial if the origin and growth of the outer Galactic 
disk were the products of infalling systems/material from the extended environment of the Galaxy over Gyr timescales \citep{mag09}. 

Looking in the opposite direction, cluster studies toward the Galactic center have been hampered by the observational limits imposed by
field crowding, high reddening, and a real deficit of surviving clusters interior to a galactocentric radius of 7 kpc. Resulting interpretations of the 
interior galactic gradient often have been affected by the presence within the sample of perhaps the most anomalous open cluster studied to
date, NGC 6791. The extreme nature of the cluster derives from the paradoxical combination of high metallicity and high age, though the
degree of the anomaly has remained a point of controversy for decades. A consensus from a variety of studies is that the cluster has an overall 
metallicity of [Fe/H] = +0.4 $\pm$ 0.1 based upon intermediate-band photometry \citep{atm07}, high resolution spectroscopy of stars in 
various parts of the CMD, including a horizontal branch star \citep{pg98}, four red giant clump stars \citep{gr06, car07}, 
ten red giant stars \citep{car06}, and two turn-off stars \citep{boe09}, high resolution infrared spectroscopy of six M giants \citep{or06},
and medium resolution spectroscopy of 24 giants \citep{wo03}.  Using three different isochrone sets \citep{de04, van06, pie04}, 
an age of 8 $\pm$ 1 Gyr has been determined from isochrone fitting of the cluster turnoff and from the mass-radius diagram of an eclipsing binary 
\citep{bed08, gru08}, though \citet{bed08} derive the puzzling age of 4-6 Gyr from the cooling curve for the cluster white dwarfs.

In most models of galactic chemical evolution, attainment of such a high degree of chemical enrichment within such a short time interval 
after the formation of the disk presents a significant challenge. Some solutions to the problem have been based upon the unusual kinematics 
of the cluster, implying that the cluster is either an interloper from a region much closer to the galactic center where a steep 
linear gradient would indicate the existence of clusters of unusually high metallicity or a cluster captured in a merger event \citep{car06}. 
Both interpretations are built upon questionable assumptions tied to the chemical composition of the cluster parent population, particularly
the latter solution requiring a super-metal-rich cluster from a dwarf system; in fact, dwarf galaxies are metal-poor \citep{sk,vz}.
As long as NGC 6791 represented a unique case, both viewpoints retained some plausibility, but with the discovery of NGC 6253, the picture is significantly altered.

The first comprehensive broad-band photometric study of NGC 6253 was carried out by \citet{bra97}. Based upon multicolor comparisons to the
morphology and luminosity functions of theoretical isochrones, \citet{bra97} concluded that the cluster was $\sim$ 3 Gyr old with a metallicity
nearly twice solar. From broad-band photometry and integrated spectra, \citet{pia98} found [Fe/H] = +0.2 and an age of 5 $\pm$ 1.5 Gyr. High-dispersion
spectra of four giants by \citet{car00} indicated [Fe/H] = +0.36 $\pm$ 0.15, a result used by \citet{sag01} to obtain an age of 2.5 $\pm$ 0.6 Gyr.
A weakness in all these discussions was the exact value of the reddening, which ranged from E$(B-V)$ = 0.20 to 0.32. The first direct measurement
of the cluster reddening and metallicity tied to intermediate-band photometry of a large sample of cluster turnoff stars \citep{tatd} 
generated E$(B-V)$ = 0.26 $\pm$ 0.003, where the error cited is the standard error of the mean (sem), and identified NGC 6253 as 
the most metal-rich open cluster known, with a metallicity at least as large 
as NGC 6791 and potentially much larger.  A best-fit approach to a match with theoretical isochrones indicated that the cluster CMD 
morphology was ideally reproduced with an age of 3 $\pm$ 0.5 Gyrs and alpha-enhanced isochrones. The weak point in the analysis was the extreme 
nature of the photometric indices which placed the cluster well beyond the limits of the photometric calibration based upon stars in the solar 
neighborhood. High-dispersion spectroscopic analysis of 4 clump giants by \citet{car07} has generated [Fe/H] = +0.46 $\pm$ 0.03, virtually 
identical to NGC 6791, while \citet{ses07} find [Fe/H] = +0.36 $\pm$ 0.07 from 4 probable members. A reanalysis of the 
intermediate-band photometry of \citet{tatd} using an improved field star calibration \citep{atm07} demonstrated that, independent of the 
adopted reddening, [Fe/H] for NGC 6253 is at least as high as that for NGC 6791 ([Fe/H] = +0.45) and could be 0.1 dex higher. The most recent
addition to the cluster database has been supplied by \citet{mon09}, who used a CCD mosaic to measure proper-motion memberships for stars in a large
area around the cluster, as well as supplying a new set of $BVRIJHK$ photometry. Adopting a low reddening (E$(B-V)$ = 0.15) from multicolor
fits to theoretical isochrones, \citet{mon09} find a slightly older age (3.5 Gyr) for the cluster than commonly derived, a possible byproduct of
comparisons with isochrones that may be too metal-poor ([Fe/H] = +0.22 and +0.36). 

While it is clear that, from a chemical composition standpoint, NGC 6253 is a younger cousin of NGC 6791, its extreme characteristics in relation to
the rest of the open cluster population and a desire to place it within the context of Galactic chemical evolution make it an important object for 
study at higher resolution, beyond the handful of clump giants and turnoff stars that have been discussed to date. For this reason, HYDRA 
spectra of 89 potential cluster members from the tip of the giant branch to below the main sequence turnoff were obtained in 
multiple wavelength regions for detailed elemental analysis. In this first of three papers we present a discussion of the results for elements other than Li 
based upon absorption lines in the region of the Li 6708 \AA\ line. Future papers will discuss other wavelength regions \citep{mad10}, as well 
as the pattern found for the Li abundance as a function of mass and evolutionary state within the cluster \citep{cum10}.

The layout of the paper is as follows: Sec. 2 presents the observations and their reduction, identification of probable members 
based upon the derived radial velocities coupled with proper-motion membership, and discussion of the photometric system used
to define the stellar colors. Sec. 3 details the determination of the fundamental stellar parameters used in the derivation of 
the abundances from the absorption lines, Sec. 4 lays out the elemental abundance analysis, and Sec. 5 summarizes our conclusions.

\section{SPECTROSCOPIC DATA}
\subsection{Observations and Reductions}
Spectra of 89 stars in NGC 6253 were obtained in 2005, 2006 and 2007 with the Cerro Tololo Inter-American Observatory Blanco 
4m telescope equipped with the Hydra multi-object spectrograph. Two distinct configurations were employed in 2005, one (RG/BS) targeting the 
red giants and blue stragglers of the cluster field, the other (MS/TO) aimed at stars at or below the cluster turnoff. In the Li 
6708 \AA\  region, the RG/BS configuration stars received a total of two hours of exposure, all on 19 July 2005. Observations for the fainter, 
MS/TO configuration stars were obtained in all three years, with the greatest portion of the total signal collected in 2007 due to the poor 
weather conditions in 2005 and 2006. For the Li region under discussion in this paper, exposures for the MS/TO configuration included 4.4 hours 
on 18 July, 2005, 12.1 hours on 31 May and 1 June, 2006, and finally 18.5 hours on 25 and 26 May, 2007. 

The spectra are characterized by a dispersion of 0.15 \AA\ per pixel in the wavelength range from 6520 \AA\ to 6810 \AA\ with S/N per pixel that ranges 
from 22 to 127 for the 35 RG/BS stars (2005 data only) and from 30 to 170 for the 54 MS/TO stars.  Our signal-to-noise measurements were 
made by sampling spectra in several spectral regions: from 6680-6690 \AA, 6646-6660 \AA\ and 6615-6622 \AA. Measurements of S/N per pixel in the
6680-6695 \AA\ region nearer the Li 6708 \AA\ line are higher and will be used in the analysis of the lithium abundances by \citet{cum10}. Measured widths 
for calibration lamp lines from 2005 and 2006 imply a spectral resolution of $15,000$ (= 3.0 pixels) in the region near 6700 \AA. 

Neither configuration utilized the full complement of 132 fibers of the Hydra instrument, leaving dozens of fibers available for measurements of the
sky background within the cluster. In practice, low-throughput fibers and the constraints of pre-assigning stellar positions in a 
crowded field limited the number of fibers usable for sky measurement to approximately 30 fibers, which still provides a significant 
advantage in determining the local background signal. Prior to the 2007 run, the ThAr lamp became contaminated and unusable; for this 
set of data, daytime sky spectra were used as spectral wavelength calibration standards.

The data were reduced using standard reduction routines in IRAF\footnote[1]{IRAF is distributed by the National Optical Astronomy 
Observatories, which are operated by the Association of Universities for Research in Astronomy, Inc., under cooperative 
agreement with the National Science 
Foundation.}  These routines included, in order of application, bias subtraction, division by the averaged flat field, dispersion correction 
through interpolation of the comparison spectra for data from 2005 and 2006, throughput correction for individual fibers, and 
continuum normalization. After flat field division and before the dispersion correction, the long-exposure program data were cleaned 
of cosmic rays using ``L. A. Cosmic''\footnote[2]{http://www.astro.yale.edu/dokkum/lacosmic/, an IRAF script developed 
by P. van Dokkum (van Dokkum 2001; spectroscopic version.}.
Repeating the procedure for stars observed in NGC 3680 from the 2005 run \citep{atd09}, spectra for each star from each night 
for a given run were combined to produce an annual dataset; a final additive combination of all frames from all years was then generated 
for the measurement of equivalent widths and the spectral synthesis comparisons.

As mentioned above, due to contamination problems with the normal comparison source, the dispersion corrections for the 2007 
spectra were based upon solar/sky spectra taken each afternoon. Comparisons of the final wavelength-defined 
spectra for NGC 6253 from 2005 and 2006 with the data set from 2007, prior to generating summed spectra from all three runs, 
proved that the spectra from all three runs were on the same wavelength and zeropoint scale within the resolution-defined uncertainties, i.e. 
the use of the solar spectra for comparison purposes in 2007 had no significant impact on the radial velocities or the equivalent 
widths. Moreover, as demonstrated by the test case of NGC 3680 \citep{atd09}, our procedures generate equivalent widths and radial velocities 
that are an excellent match to previous high dispersion studies when a statistically significant sample overlap is available.

\subsection{Basic Data: Radial Velocities and Cluster Membership} 
Because it is the only identification set common to all stars in our sample, stars will be referred to using 
NGC 6253 identifications on the system of the open cluster database, WEBDA\footnote[3] 
{http://www.univie.ac.at/webda}. Radial-velocity standards were observed on each night of every run 
to provide an external check on the wavelength calibration and the velocity scale. Individual stellar radial 
velocities were derived from each 
annual composite dataset utilizing the fourier-transform, cross-correlation facility FXCOR in IRAF. In this utility, program stars are compared to 
stellar templates of similar effective temperature ($T_{eff}$) over the wavelength range from 6575 \AA\ to 6790 \AA, as well as a narrower region in the vicinity of 
H$\alpha$ alone. We have also included in our analysis the radial-velocity measurements obtained in 2007 from spectra in the region of the 
oxygen triplet \citep{mad10}. These radial-velocity measurements were based exclusively on measured wavelengths for lines in the daytime solar 
spectrum as well as the spectra for program stars.

Even though the 2005 and 2006 data contribute relatively little to the S/N of the combined spectra for the MS/TO stars, 
these datasets do provide the potential for highlighting radial-velocity variations. With radial velocity ($V_r$) and rotational velocity 
information in hand for each annual dataset, our goal was to combine the datasets to produce radial velocities of high internal precision, 
and then establish the accuracy of the radial-velocity scale with respect to external radial-velocity standards.  
The first step ensures not only average $V_r$ values of high precision but permits the identification of variable velocity 
(SB) candidates. The radial-velocity datasets based on spectra in the Li region for 2005 and 2007, as well as the additional dataset 
from 2007 covering the spectral region of the oxygen triplet, were mapped to the radial velocities of the 2006 dataset in the
Li region by determining, then removing, average differences between individual radial velocities in each dataset and the 
2006 radial velocities. The four datasets on a common radial-velocity system were combined and averaged; the dispersions 
associated with the averaged velocities are independent of any uncertainty about the absolute scale of radial velocities, i.e.
they remain the same whether one adopts 2006 or any other dataset as the standard. 

The immediate benefit is that five candidate spectroscopic binaries with standard deviations above 10 km/sec are easily picked out of 
the sample of 54 MS/TO stars, as well as seven MS/TO stars with radial velocities that differ from the eventual cluster average
by 7 km/sec or more, equivalent to more than five standard deviations, i.e. possible non-members 
or members within a long-period binary. If these probable SB and/or possible non-members from radial-velocity data alone are excluded, 
the dispersion among the averaged mean radial velocities for the remaining 42 stars is $\sim 1.2$ km/sec, quite close to the 
average error for a single FXCOR measurement, indicating that the averaging process has not introduced significant scatter 
into the average radial velocities through the inclusion of unrecognized non-members or candidate SB's.

Comparison of the derived and published velocities for our radial-velocity standards suggested the need for a correction to this 
preliminary average velocity scale; this correction of -1.27 km/sec introduces an added uncertainty in the zero-point of the radial-velocity scale 
of $\pm 0.26$ km/sec. The added component is included in the standard deviation associated with the average radial velocity for 
cluster MS/TO members of -29.32 $\pm$ 1.30 km/sec or $\pm$ 0.2 km/sec (sem). Note that, as discussed below, two of 
the MS/TO stars classed as members through their radial velocities are probable proper-motion non-members. If these two 
stars (7638, 7806) with $V_r$ = -30.23 and -28.19 km/sec are excluded, the average cluster radial velocity remains unchanged. We will 
return to the final membership classification using the combined insight from both radial velocities and proper-motions after discussing
the RG/BS stars.

A single dataset provides information about the radial velocities for the red giant and blue straggler (RG/BS) candidates observed in 2005. 
No identification of radial-velocity variability is possible; the radial-velocity scale was adjusted based on comparison to the appropriate 
night's radial-velocity standards. For the subset of 18 apparent radial-velocity members, the average cluster velocity 
is -29.60 $\pm$ 1.50 km/sec (sd). Three of the stars excluded from this average (5201, 2542, 2126) are probable members from 
proper-motion data. One possible blue straggler (5201) has a larger than average uncertainty in $V_r$; stars 2542 and 2126 
have $V_r$ = -24.52 and -37.14 km/sec, respectively, so their inclusion in the average for
the evolved stars would lower the mean velocity by 0.12 km/sec. Fig. 1 shows the histogram of radial velocities for the combined RG/BS and 
MS/TO samples. For the latter, probable binaries have been excluded from the counts; the velocity range used to define cluster membership 
from radial velocity alone is noted.  The average of all probable radial-velocity single members, MS/TO as well as evolved stars, is 
-29.41 $\pm$ 1.31 km/sec (sd) or $\pm$0.17 (sem).
 
We can compare our cluster velocity results with two recent studies of NGC 6253. \citet{car07} studied five red giant stars. One of 
the five has a velocity of -20.6 km/sec; removing it from the sample produces a tighter average cluster velocity 
of -28.26 $\pm$ 0.6 km/sec (sd)
among the remaining four giants.  A similar result is obtained by \citet{ses07} from seven stars ranging from the turnoff to the 
red giant branch 
clump. A rather noisy average suggests that not all are single and/or member stars; the average of the four most similar velocity values is 
-29.71$ \pm$ 0.79 km/sec (sd). 

More specific, star-by-star comparisons are also possible with respect to both of these spectroscopic studies. Three of the 
five stars studied 
by \citet{car07} are within our sample of RG/BS stars. Our velocities are, on average, $0.35 \pm 0.46$ km/sec lower than those of \citet{car07} 
for stars 3595, 2885 and 4510, i.e. not significantly different. Two of the stars studied by \citet{ses07} are within our survey sample, 
although one of them (2542) is considered a likely non-member in both studies. For the other star, 3138, our velocity is 0.6 km/sec lower, 
within the individual errors of both surveys. 

Of the 89 stars in our sample, 41 have proper-motion membership estimates from \citet{mon09}. A plot of the radial-velocity measures, 
with error bars, for these 41 stars as a function of proper-motion probability is illustrated in Fig. 2. For the final
membership classification, any star with a radial velocity that differed from the cluster mean by more than 5.2 km/sec (four
standard deviations) was classed as a radial-velocity non-member. Two stars with velocities offset by just under 5.2 km/sec (2542, 5201) 
are tagged as possible radial-velocity nonmembers. Keeping in mind that proper-motion data are available for only 41 stars in our
sample, any star with a probability below 50\% was classified as a definite proper-motion non-member. Combining the results from both 
membership criteria whenever possible, the agreement between the two indicators is excellent. Excluding probable binaries with
large radial-velocity dispersions, only four stars are classed as having uncertain final membership, 2542, 5201, 2126, and 7348.
Our final designation of membership based upon the combined membership classification is represented in Fig. 2 by circles (probable members), 
squares (probable non-members), and triangles (uncertain membership).  

Table 1 presents the WEBDA identifications, radial-velocity measures, $BV$ photometry from \citet{tatd} and \citet{mon09}, 
membership based upon radial velocity alone, spectroscopic signal-to-noise per pixel values, proper-motion membership probabilities, 
and a final overall 
determination of membership for the 89 stars observed in NGC 6253. Where multi-year observations were available, the dispersion from 
year-to-year radial velocities is indicated in the error column; for the candidate RG/BS stars, the radial-velocity error from a single 
observation is indicated.

\subsection{Basic Data: Photometry}
Photometry from \citet{tatd} has been used as the primary source for $B-V$; the reader is referred to this paper for details on the 
compilation and merger of the published CCD photometry used to construct the $BV$ catalog. Fig. 3 illustrates the CMD distribution of
the stellar sample with filled symbols identifying probable member binary-star candidates. Squares and circles designate stars observed within
the RG/BS and MS/TO categorization, respectively; crosses and asterisks are the probable non-members among these two respective categories.
Four triangles designate the stars with uncertain membership.

It is important to emphasize that the majority of the MS/TO star candidates were selected for inclusion in the spectroscopic survey primarily on the 
basis of their identification as probable members through their locations in the CMD, coupled with analysis of the 
intermediate-band photometry of \citet{tatd}. Of the 52 stars tagged as possible members through the photometry, 44 ended up as 
probable members from both radial-velocities and proper-motions, with only 5 stars classed as likely binaries, an exceptional success rate that 
indicates the efficiency of the intermediate-band approach for optimizing sample selection among fainter stars in clusters when alternative 
membership information is lacking.

The $B-V$ data of \citet{tatd} were compiled by converting the observed $b-y$ CCD indices into $B-V$ through a direct comparison 
to the merged CCD $B-V$ data from three independent sources \citep{bra97, pia98, sag01}. This approach was taken because
the internal errors of the $b-y$ photometry were considerably smaller than most of the quoted errors for the published 
$BV$ surveys and the $uvby$ study covered a significantly larger field. The average standard errors of the mean for $V$ and $b-y$
for the MS/TO sample are 0.0033 and 0.0048, respectively. The comparable data for the RG/BS stars are 0.0031 and 0.0043, respectively. 
Transferred into $B-V$, the average standard errors of the mean in color become 0.0083 for MS/TO stars and 0.0063 for the RG/BS stars.
The system of \citet{sag01} was selected as
the standard because it agreed with that of \citet{pia98}, though with better internal errors, and both of these studies
showed no color dependence among the residuals in $V$ when compared to the $uvby$ value of $V$. By contrast, the photometry of
\citet{bra97} exhibited a significant color term in $B$ when compared to the other sources and a smaller but real color term in 
$V$ when compared to the $V$ magnitudes from the $uvby$ data. With the CCD mosaic study of NGC 6253 by \citet{mon09} now available,
how do the photometric systems compare? 

The data from \citet{mon09} were cross-matched with the stars observed by \citet{tatd} and the residuals in $V$ determined, in the sense
(MON - TAD), for approximately 2400 stars brighter than $V$ = 18.0. The patterns that emerged are best illustrated using the stars brighter
than $V$ = 16.5 to minimize confusion caused by the larger sample size and scatter at fainter magnitudes. Fig. 4a shows the residuals 
as a function of the X-pixel coordinate of the Wide Field Imager, a mosaic of eight 2K x 4K CCDs. (For details regarding the 
instrumentation, the reader is referred to the discussion by \citet{mon09}.) The boundaries of three of the CCD chips within the mosaic 
array that overlap the $uvby$ field in the X-pixel direction are easily identified, as is the discontinuity in residuals for stars on the 
chip located at X greater than 610 pixels. From 758 stars 
with X below 610 pixels, the average residual in $V$ is +0.009 $\pm$ 0.014 mag (sd); for 168 stars above X = 610 pixels, the mean residual 
is -0.034 $\pm$ 0.014 mag. Fig. 4b shows the comparable info for the CCD Y-pixel coordinate direction, where each chip 
identified in X covers the entire 
Y direction without a break. The offset created by the pattern in X is obvious. Besides this, there is a small trend in the residuals with Y. 
Between Y = -1000 and +900 pixels, the residuals shift by $\sim$0.02 mag. Whether this arises from the $uvby$ data, the $BV$ data, or both 
cannot be determined. This level of change is of little consequence and the offset in $V$ at large X is readily corrected,
if necessary. The small scatter among the residuals is gratifying and it appears that the $V$ photometry is on the same system 
at the $\pm$ 0.01 mag level. Moreover, if we assume that the dispersion in the $V$ residuals is due to equal contributions by both
sources of photometry, the upper limit on the errors in $V$ for the data adopted in this study becomes 0.010 mag.  
However, we emphasize that for the purposes of this investigation, use of the $BV$ system of \citet{tatd} remains
unchanged; $V$ magnitudes and the color-based temperatures discussed in Sec. 3 do not include the photometric data of \citet{mon09}.

For the current investigation, the critical question is the $B$ photometry or, equivalently, $B-V$. Unlike $V$, we cannot use a direct comparison from
$uvby$ data to test the reliability of the $B$ zero-point or color dependence. We can, however, test for consistency among the four CCD
surveys now available. Fig. 5a shows the residuals in $B-V$, in the sense (MON - REF), as a function of $(B-V)_{MON}$ for the photometry of 
\citet{pia98} (crosses) and \citet{sag01} (open circles). As before, we have restricted the samples to the brighter stars to illustrate 
the pattern. Confirming the analysis in \citet{tatd}, the color dependence of the residuals is the same for both surveys, implying that 
the \citet{mon09} data are too blue at the cluster turnoff ($(B-V)$ = 0.8) by $\sim$0.04 mag, with the offset growing to 0.12 mag by $B-V$ = 1.3.
The analogous comparison for the data of \citet{bra97} is shown in Fig. 5b. At the cluster turnoff, the zero-point offset found in Fig. 5a
is confirmed. However, the steep color dependence among the residuals is dramatically reduced. This presents us with two possible zero-points for
the $B-V$ colors, as defined by the color of the stars at the cluster turnoff, and two options for the differential color of the giant branch
relative to the turnoff. For this investigation, we will adopt the $BV$ data of \citet{tatd} as our photometric system and explore
the impact of the alternate choices on our conclusions. In general, a bluer zero-point and/or a smaller differential color between the giant branch
and the turnoff will increase our metallicity estimates for the turnoff stars and/or the giants.

\section{FUNDAMENTAL STELLAR PARAMETERS}
The reliability of the final stellar elemental abundances is dependent in large part upon the accuracy of the stellar parameters that 
characterize the model atmospheres used in interpreting the measured equivalent widths. Once membership is established, the three primary
parameters of interest are the effective temperature, surface gravity, and microturbulent velocity, with different sensitivities 
to each for giants and dwarfs.

\subsection{Reddening and Photometric Temperatures}
What is the foreground reddening to NGC 6253? Can we predict the unreddened colors of highly metal-rich stars? These are separate questions with 
unfortunately intertwined answers, both of which bear upon the larger and more critical issue of effective temperatures based upon photometric colors. Since
reddening estimates are often tied to color comparisons to unreddened stars of similar metallicity, it is imperative that a statistically
significant sample of minimally reddened field stars be accessible for comparison. At the expected metallicity and effective temperature range of NGC 6253, 
nearby field stars of appropriate effective temperature are few and far between, requiring the extrapolation of trends from a lower [Fe/H] range. 
However, there is reason to suspect that metallicities well above solar affect optical colors by amounts that exceed the normal extrapolations, 
a point we will return to below. 

A related issue worth considering is whether a helium abundance significantly different from a near-solar value could, by itself, alter 
the stellar colors appreciably. \citet{gir07} explored this possibility with an explicit reassessment of the effects of helium abundance 
differences on bolometric corrections and color-temperature relations. Their approach included a recomputation of ATLAS9 models for a range 
of compositions spanning increments of 0.1 in Y, the helium abundance by mass fraction, and reaching [Fe/H] values significantly above solar. 
Based upon their conclusions, it seems very unlikely that even extreme helium abundance differences would appreciably affect the $(B-V)$ 
colors for stars approaching the Sun's surface gravity and effective temperature.

To constrain the cluster reddening while sidestepping the question of the $BV$ photometry, a number of options exist. First, an upper limit 
on the foreground reddening to NGC 6253 may be set via the dust maps of \citet{sch98} in the direction of the cluster; E$(B-V)$ = 0.35 is 
found for NGC 6253's galactic coordinates. However, as discussed in previous papers, given the low galactic latitude of the cluster and 
the correlated modest distance away from the plane ($\sim$ 190 pc), NGC 6253 is probably affected by less than this full amount, although
the exact fraction is unknown. Moreover, \citet{arc99} indicate that the reddening maps of \citet{sch98} overestimate the size of the reddening
for estimates above E$(B-V)$ = 0.15. If we adopt the relation for adjusting the high-end reddening as determined by \citet{ka06}, the 
\citet{sch98} value for the direction of NGC 6253 is reduced to E$(B-V)$ = 0.26.

Second, one can use intermediate-band, $uvbyCa$H$\beta$ photometry to define the intrinsic colors of the stars near the turnoff 
under the assumption that the line index, H$\beta$, is metallicity and reddening-independent. Using traditional intrinsic-color 
relations tied to stars in the solar neighborhood, \citet{tatd} found E$(B-V)$ = 0.26 $\pm$ 0.003. Ongoing investigation of the 
fundamental calibrations of the photometric system for cooler and/or more metal-rich stars \citep{att02, tat02, tatf, tvat} has 
demonstrated that, when projected to very high metallicity, the standard intrinsic color relations for cooler dwarfs tend to underestimate 
the value of the $b-y$ index, leading to an overestimate of the reddening. This implies that E$(B-V)$ = 0.26 should be a plausible 
upper limit for the true reddening value. The size of the metallicity effect has been forcefully demonstrated for NGC 6791 \citep{atm07}, 
where the traditional calibration from intermediate-band analysis led to E$(B-V)$ = 0.23, well above the limit imposed by the dust 
maps of \citet{sch98}. Direct comparison with field stars of comparable effective temperature and metallicity produced a more plausible E$(B-V)$ 
= 0.15-0.16.  An attempt to correct the comparable problem for NGC 6253 proved less satisfactory due to the shortage of hotter dwarfs 
with spectroscopic abundances and intermediate-band photometry at the metallicity derived for NGC 6253. Using an extrapolation from 
stars slightly more metal-rich than the Hyades, \citet{atm07} found E$(B-V)$ = 0.16, the same value as NGC 6791, for [Fe/H] = +0.57, 
on a scale where NGC 6791 has [Fe/H] = +0.45. If the metallicity of NGC 6253 is closer to that of NGC 6791, the reddening
must be higher. We regard the E$(B-V)$ = 0.16 estimate as a lower limit to the true cluster value as defined by intermediate-band photometry.

A CMD-based third option is to make use of broad-band IR photometry of giants in the two clusters of comparable metallicity, NGC 6791 and
NGC 6253. Following the approach of \citet{car05}, we collected $JHK$ magnitudes and colors with the highest quality index from the 
2MASS data archive for stars near NGC 6253. Matching the 2MASS positions to the existing photometric survey effectively restricted this search to stars 
within 6.5\arcmin\ of the cluster center. Figure 6 shows an overplot of the $K, (J-K)$ color-magnitude diagram for NGC 6791 (triangles) and 
NGC 6253 (circles). A color offset of $+0.06$ mag has been applied to the $(J-K)$ colors for the NGC 6791 stars, while a 2.1 mag adjustment 
has been applied to the fainter NGC 6791 $K$ magnitudes. The alignment of the giant branches and the red clumps of the two clusters suggests 
that the foreground reddening, E$(J-K)$, for NGC 6253 is at least 0.06 mag larger than that of NGC 6791. The conclusion that this value 
represents a lower limit results from a presumption that NGC 6253's younger age should result in an intrinsically bluer giant branch with 
respect to that of NGC 6791. However, the alignment in the $JK$ CMD is set by the assertion that the absolute magnitude and color of the 
clumps are the same in $K$ and $(J-K)_0$, independent of age, at least for older clusters. For a complete discussion of this important issue, 
see \citet{car05}. Following \citet{car05}, we employ reddening relations from \citet{rie85} to convert $\Delta$E$(J-K) \geq 0.06$ to 
$\Delta$E$(B-V) \geq 0.12$ with respect to NGC 6791. Using $JHK$ analysis, \citet{car05} cite a result of E$(B-V) = 0.14$ for NGC 6791, 
implying a reddening of at least E$(B-V)$ = 0.26 for NGC 6253. As a compromise among the range of potential values, E$(B-V) = 0.22$ $\pm$ 0.04 
has been used to correct the NGC 6253 stars for reddening. 

For effective temperature estimates for the MS/TO stars and the blue stragglers, we have maintained use of the color-temperature relation 
developed by Deliyannis and co-workers, as detailed in \citet{atd09}. We have also compared effective temperatures derived from $(B-V)$ colors to 
similar results based upon use of the \citet{ram05} color-temperature relations.  For both color-temperature scales, if [Fe/H] $ = 0.45$ 
is assumed as characteristic of the cluster stars, the scales are essentially indistinguishable, differing by less than 10 K over 
the applicable and narrow range of effective temperatures for the turnoff and main sequence stars. For the evolved stars, we converted $(B-V)_{0}$ 
to T$_{eff}$ using the color-temperature relation of \citet{ram05}, a revision of the dwarf and giant calibrations of \citet{alo96, alo99}. 
For the effective temperature and gravity range of our red giant stars, the color-temperature relations of \citet{ram05} produce 
results essentially identical to the slightly older relations of  \citet{alo96, alo99}; they are explicitly designed to be applicable 
to stars as metal-rich as [Fe/H] = 0.5.

As a test of the consistency of the effective temperature scale for the turnoff stars implied by the adoption of $E(B-V) = 0.22$ and the
$B-V$ zero-point of the photometry excluding the data of \citet{mon09}, we appealed to the high-dispersion spectroscopic catalog of field stars by 
\citet{val05}, expanded and enhanced in \citet{tvat,tate}, to compare H$\beta$ indices and spectroscopic temperature estimates for the small 
sample of catalog stars with [Fe/H] $ \geq 0.30$. This relationship was applied to the normal sequence defined by turnoff and main sequence 
stars in NGC 6253 to map effective temperatures to observed $(b-y)$ colors. This mapping relation predicts that turnoff stars characterized by 
$(b-y) = 0.57 \pm 0.01 $ should have effective temperatures of 5935 $\pm$ 60 K. This corresponds precisely with the predicted effective temperatures for 
these stars using the adopted color-temperature calibration with a reddening value of E$(B-V)$ = 0.23. Since we have adopted a correction 
of 0.22 mag to the $(B-V)$ colors in NGC 6253, use of higher reddening would lead to higher derived elemental abundances. 
Correspondingly, from 34 turnoff members between $B-V$ = 0.82 and 0.93, the average difference in $B-V$, in the sense (Table 1 - MON), 
is 0.039 $\pm$ 0.017 mag. Adoption of the \citet{mon09} zero-point for the $BV$ photometry is equivalent to raising the implied reddening 
for the analysis of stars at the turnoff on our adopted photometric zero-point to E$(B-V)$ = 0.26, generating higher abundances for the stars 
at the turnoff.

Since the colors are used predominantly for the purpose of setting the effective temperature scale, the real question is whether the color shift resulting 
from the combined effect of the reddening correction and the $B-V$ zero-point leads to correct effective temperatures. If it 
were possible to derive the
effective temperatures from spectroscopic line analysis, independent of the colors and the reddening, errors in the photometric zero-point would 
translate into compensating errors in the derived reddening, leaving the color temperature the same as that predicted by the spectroscopic 
analysis, although in that case, additional errors would be introduced due to the greater dependence  of the analysis on the 
model atmospheres. For reasons discussed below, this approach was only possible for two stars at 
the turnoff; the derived spectroscopic temperatures were in excellent agreement with those produced from the color calibration with
the photometry and reddening as adopted.

How does our effective temperature scale compare with that of previous spectroscopic studies? Turnoff stars have only 
been observed by \citet{ses07} and neither of their two turnoff stars overlap with our study. However, because our $B-V$ colors for 
the vertical portion of the turnoff are on the same color system as \citet{bra97}, we can compare the temperature scales as 
predicted from our photometry with the photometric scale derived by \citet{ses07}. Keeping in mind 
that \citet{ses07} adopt E$(B-V)$ = 0.23, the average photometric temperature for stars 3053 and 2225 is 5937 K using, 
in their case, the $(B-V)$ - $T_{eff}$ relation for dwarfs from \citet{alo96}. Our average photometric temperature for the 
same stars with our adopted value of E$(B-V)$ = 0.22 is 6011 K. Since we are using the same color indices, the offset is 
entirely a result of the slight change in reddening, our higher adopted value of [Fe/H] within the color-$T_{eff}$ relation, 
and use of the revised color-$T_{eff}$ relation. More important from the standpoint of the final abundance determinations, the 
average effective temperature for these two stars derived by \citet{ses07} from spectroscopy, independent of the reddening 
and used in the abundance analysis by \citet{ses07}, is 6125 K. \citet{ses07} estimate an upper limit to the uncertainty in 
their spectroscopic temperatures of $\pm$70 K. Therefore, were we to derive our abundances for the turnoff stars using the 
spectroscopic effective temperature scale of \citet{ses07}, our abundances would increase.  The exact size of the change will 
be dealt with in Sec. 4.

For the giants, our samples overlap with two stars, 2542 and 3138. As one might expect given that our $B-V$ system for the giants 
is 0.065 mag redder and that we adopted a reddening value that is 0.01 mag smaller, the difference in the photometric temperatures, 
in the sense (Table 4 - SES), averages -80 $\pm$ 14 K for \citet{ses07}, i.e., our temperatures are cooler. 
The analogous offset with the spectroscopically derived effective temperatures is -34 $\pm$ 6 K, implying that our 
current abundances for the giants are only slightly less than or equal to what they would be if we had used 
the spectroscopic effective temperature scale of \citet{ses07}.

By contrast, the effective temperatures of \citet{car07} are purely photometric and are based upon $V-K$, bypassing 
the concern raised by the $B$
photometry. Adopting E$(B-V)$ = 0.23, the average difference in the photometric temperatures, in the sense (Table 4 - CAR), from three 
giants in common is 147 $\pm$ 34 K, implying that, relative to \citet{car07}, the spectroscopic and photometric scales of \citet{ses07} are 
too hot by 180 K and 227 K, respectively; the estimated uncertainty in the temperatures of the giants as derived by \citet{car07} is
$\pm$70 K. The temperature difference goes the wrong way in that it implies that our adopted $B-V$ colors are too blue. 
Obviously, if the abundances of \citet{car07} were derived using our photometric temperature
scale or either effective temperature scale of \citet{ses07}, their metallicities for the giants would increase, 
but exactly how much depends upon any correlated effects of altering the microturbulent velocity and the surface gravity, 
a point we will return to in Sec. 4.

\subsection{Surface Gravity and Microturbulent Velocity}
Another necessary input parameter for our spectroscopic analysis is a surface gravity estimate for each star. Without an extensive list 
of measured lines of different ionization states, we were unable to determine log $g$ from the spectra alone. Instead we tested the isochrones 
of Chaboyer and collaborators \citep{des08} and the Yale-Yonsei ($Y^2$) isochrones \citep{de04} to supply an appropriate set for a match to the 
abundance and age of NGC 6253. Scaled-solar isochrones for [Fe/H] = 0.45 were obtained from the web interface for the Dartmouth Stellar 
Evolution (DSE) database (stellar.dartmouth.edu/$\sim$models/index.html) and the $Y^2$ web site (http://www.astro.yale.edu/demarque/yyiso.html). 
With [Fe/H] and E$(B-V)$ fixed, only the distance modulus remains as a free parameter. Optimal matches were determined by eye, with special emphasis
on the turnoff region and subgiant branch luminosity due to the inconsistency of the colors of the isochrone giant branches for any plausible
age estimate. For DSE, an isochrone with an age of 3.3 Gyr matches the turnoff and subgiant branch of NGC 6253 
reasonably well for an assumed reddening E$(B-V)=0.22$ and apparent distance modulus of $(m-M)$ = 12.0, but fails to reproduce the colors and 
magnitude of the giant branch and giant branch clump; more specifically, while the isochrone values for log $g$ for the clump giants 
correspond to an expected value of $\sim 2.5$, the isochrone effective temperatures are generally 200 K cooler than indicated by photometric 
color-temperature relations. By contrast, the $Y^2$ isochrones with an age of 3.0 Gyr and an apparent modulus of 11.9 provide the optimal match
to the cluster with E$(B-V)$ = 0.22. The predicted giant branch is also redder than the cluster, but by a significantly smaller amount than
the DSE comparison. The comparison for $Y^2$ is illustrated in Figure 7, in which scaled-solar isochrones with [Fe/H]$ = 0.45$ have 
been adjusted by an apparent distance modulus of 11.9 and E$(B-V) =0.22$ and compared to photometry compiled by \citet{tatd}.  
The ages of the presented isochrones are 2.6, 3.0 and 3.4 Gyr.  

A simple match of $V$ magnitudes to log $g$ values from the fitted isochrones sufficed for main sequence, turnoff and subgiant branch 
stars; the two independent isochrone sets agreed in log $g$ at the $\pm$0.03 level, a difference too small to impact the final abundances. 
Because of the failure of both sets of isochrones to superpose upon the observed giant branch, the giant branch stars clearly needed a 
different approach. In the DSE isochrones for this high metallicity, the surface gravity of the clump stars increases gradually 
from log $g$ = 2.5 at ages below 3 Gyr to values near 2.6 for ages above 4.5 Gyrs. A value of 2.55 was chosen as the 
representative surface gravity for stars populating the red giant branch clump in NGC 6253. The $V$ magnitudes for the giant branch stars 
were artificially offset to force a match between the magnitude of the cluster's red giant branch clump to the comparable feature in 
the isochrones at log $g$ = 2.55, i.e. the surface gravities were derived differentially relative to the clump luminosity.

Microturbulent velocities ($V_t$) for F and G dwarfs are commonly estimated from a formula developed by \citet{edv} which depends on effective 
temperature and surface gravity.  Previous spectroscopic analyses by the Indiana group have used this formula exclusively, but its development 
did not include stars with metallicities as high as those found in NGC 6253. An ideal transformation of measured line equivalent widths into 
abundances uses the 2002 version of the local thermodynamic equilibrium (LTE) line analysis/spectrum synthesis code MOOG \citep{sne73} 
and appropriate stellar atmospheres. It is possible, in principle, to determine the optimum effective temperature and microturbulent 
velocity by minimizing the trend of derived abundances as functions of excitation potential and equivalent width for different lines. In practice, 
most of the spectra produced indeterminate results with this approach; finding ``clean" lines in spectra this metal-rich is a challenge.
For only two stars among the MS/TO stars, 4375 and 2193, was it possible to independently arrive at effective temperature and 
microturbulent velocity 
estimates.  Based upon the implied microturbulent velocities for these two stars, $V_t$ values from the \citet{edv} formulation appeared to be 
too large by 0.2 km/sec and 1.1 km/sec. For the present analysis, the intercept of the \citet{edv} relation for 
microturbulent velocities was reduced by 0.8 km/sec, a weighted average of the individual offsets, retaining 0.8 km/sec as a minimum value.   The sensitivity of [Fe/H] to microturbulent velocity implies that the derived [Fe/H] values are boosted by $\sim$0.2 dex by
this adoption of a lower $V_t$ scale in the analysis of the equivalent widths among the dwarfs.

As a modest check on our approach, we note that in their discussion of stars near the main sequence turnoff \citet{ses07} 
derived microturbulent velocities from analysis of the spectra, finding a microturbulent velocity of 1.27 km/sec for the one probable member, 3053.
The prescription from \citet{edv} would predict a microturbulent velocity nearly 1.0 km/sec higher, using their derived effective 
temperature and estimated surface gravity.

The recipe for microturbulent velocities among dwarf stars cannot be applied to giants. Based upon extensive analyses of open
cluster giants by the Indiana group using similar spectra (see, e.g. \citet{jac09} and references therein) 
we set the microturbulent velocity scale for the clump stars to a value of 1.5 km/sec. For evolved stars brighter (fainter) than 
the clump, the gravity-dependent formula of \citet{car04} was used to increment (decrement) $V_t$ through a differential comparison to 
the log $g$ of the clump stars. 

\section{ELEMENTAL ABUNDANCE DETERMINATION}
\subsection{Dwarf Abundances - Fe, Si, Ca, Ni}
The determination of elemental abundances in NGC 6253's candidate MS/TO stars began with the adoption of a list of cleanly measurable 
iron lines used in common with other cluster investigations by the authors within the Indiana University group, an approach that should 
enhance the consistency with other cluster results generated from the same procedure. The very strong lines in the spectra for NGC 6253 
members, however, made adherence to this original line list difficult. To identify alternative lines as replacements for the iron lines 
that were crowded by nearby strong lines, we synthesized predicted equivalent widths for lines in linelists generated from atomic data made 
available by  Kurucz\footnote[4]{http://cfaku5.cfa.harvard.edu/linelists.html} using 
MOOG and an atmospheric model with characteristics similar to our turnoff stars. All model atmospheres were generated from 
the \citet{ku92} model atmosphere set for specified $T_{eff}$, log $g$, microturbulent velocity and input abundance.  
Our final linelist consisted of 15 iron lines, to which three measurable lines of neutral Ni (6643, 6768 and 
6772 \AA) and one line each for Si (6721 \AA) and Ca (6717 \AA) were added. 

Equivalent widths were measured using the IRAF routine SPLOT, then processed using the force-fit abundance 
routine {\it abfind} in MOOG. An estimated error for each equivalent width measurement was made using a formula 
described by \citet{del93} using the width of the line and the S/N ratio per pixel. For a spectrum with a S/N per pixel of 100, the 
error in the equivalent width is approximately 5 m\AA\ .  
These steps were replicated for 20 separate exposures of the solar spectrum obtained from our daytime sky exposures from the 2007 run, adopting 
an effective temperature of 5770 K, surface gravity of log $g$\ = 4.40 and a microturbulent velocity of 1.14 km/sec for the Sun. 
Signal-to-noise per pixel values for these solar spectra ranged from 102 to 195. All resulting stellar abundances were differenced 
with respect to the generated solar abundances on a line-by-line basis. This minimizes the impact of the specific choice of 
log $gf$ values on the final abundances from the individual lines. All measured equivalent widths for the 38 probable member MS/TO stars are listed in Table 2, available in full form in the on-line edition of the journal.

Table 3 presents [Fe/H] estimates for each of 38 probable member dwarfs, based on 9 to 15 iron lines per star. For elements with multiple lines, 
the standard error of the mean abundance is presented; for elements with single lines, the error presented has been propagated from the formal error 
in the equivalent width measurement. Table 1 includes 40 single probable members among the MS/TO sample. The spectrum for the very 
faintest star, 7822, could not be measured with any degree of security. Another star, 2562, was included in the observing configuration 
composed of MS/TO stars but sits at the base of the red giant branch, and will be included with the red giant sample.  
Some individual equivalent width measurements were discarded in the course of analysis: 
equivalent widths less than 15 m\AA\ (approximately 3 times the typical error in equivalent width) or greater 
than 200 m\AA\ , for which a line would be considered strong, as well as the few measures with calculated formal errors 
in the equivalent width that were comparable to the EW measurement. 
 
A histogram of individual iron abundance measures was constructed for all lines and all member stars in the MS/TO sample, 
following which measures of log A(Fe) below 7.4 and above 8.5 were excluded (corresponding to [Fe/H] above 1.0 or below -0.1). All 
individual values of log A(Fe) were converted to a linear scale before averaging; the logarithmic abundance was recreated 
following an averaging 
procedure weighted by individual measurement errors. From 38 turnoff members, the weighted average abundances are found to be
$0.43 \pm 0.01$, $0.53 \pm 0.02$, $0.43 \pm 0.03$ and $0.61 \pm 0.02$ for [Fe/H], [Ni/H], [Si/H], and [Ca/H], respectively. The errors 
refer to the standard error of a weighted mean; because we averaged the numerical abundances rather than the logarithmic quantities, the 
resulting errors are asymmetric.  We note that for abundance estimates covering a realistically large range, the order of computational 
operation is critical; the consequence of averaging logarithmic abundances would be an estimated average [Fe/H] nearly 0.1 dex lower over 
our large sample of abundances from 504 iron lines. 

Figures 8 and 9 provide representative illustrations of the lack of a significant dependence of the derived iron abundances on 
$V$ magnitude and on wavelength, where the plotted abundance is the appropriate average for each MS/TO star in Figure 8, for each line in 
Figure 9. Errors shown in both figures are standard deviations. A linear fit through the data in Fig. 8 does indicate a weak dependence 
(slope of 0.08 $\pm 0.03$ with a correlation coefficient of 0.42) of the abundance on position within
the vertical turnoff with the stars at $V$ = 14.75 ending up 0.15 dex more metal-poor than stars at $V$ = 16.5. The statistical significance of the 
slope is rather marginal, making it difficult to know whether this is a real effect tied to physical processes occurring in the outer layers of stars, 
or whether this is a byproduct of our reduction procedure. Keeping this caveat in mind, one might wonder whether we have detected the effects of 
microscopic diffusion, however marginally.

\subsection{Evidence for Microscopic Diffusion?}
Microscopic diffusion can affect the elemental abundances observed on stellar surfaces.  The efficiency of convective mixing wipes out any 
diffusion that might have occurred inside a convection zone; however, gravitational settling and thermal diffusion can drain elements out of the
{\it bottom} of the surface convection zone (SCZ), and thus reduce the surface abundance.  Conversely, radiative levitation can push elements
into the SCZ, thereby enriching the surface abundances (e.g. \citet{mic86}).  The efficiency of downward diffusion generally increases toward 
the surface, so cooler dwarfs with deeper SCZs should suffer the least diffusion, while progressively hotter dwarfs (with shallower SCZs) should 
show progressively more diffusion, and stars near the turnoff should show the most.  Deepening SCZs during subgiant and giant evolution would erase 
diffusion-related surface abundance anomalies created during the main sequence, either by dredge-up of material diffused downward during the 
main sequence \citep{del90}, or by dilution of radiative levitation-enhanced turnoff surface abundances.  Some evidence exists for 
downward diffusion in turnoff stars of the very metal-poor globular cluster M92 \citep{kin98, boe98, ram01, kor07} but 
\citet{gra01} find no differences between turnoff stars and giants in the somewhat more metal-rich globulars NGC 6397 and NGC 6752, 
although \citet{kor07} claim that such differences do exist in NGC 6397. \citet{ric02} studied the metallicity dependence of diffusion in models 
spanning [Fe/H]= -4.3 to -0.7, and argued that the most metal-poor turnoffs would show the greatest amount of diffusion, due in part to the fact 
that increasing metallicity results in cooler turnoff $T_{eff}$ at a given age. Thus, it could very well be the case that only the most 
metal-poor globulars, such as M92, will show significant differences between the turnoff and the RGB.

\citet{mic04} constructed stellar evolutionary models that included diffusion for solar-metallicity stars, with particular application to the 
old open clusters M67 and NGC 188. (We are not aware of any diffusion models for super-metal-rich stars such as those of NGC
6253.) They assumed turnoff $T_{eff}$ of about 6150 K and 5950 K, respectively, which are close to our turnoff $T_{eff}$ for NGC 6253 of 
about 6000 K; recall also that NGC 6253 could easily be a little hotter or cooler (Sec. 3.1). Near the turnoff, the degree of diffusion 
becomes a steep function of $T_{eff}$: for example, for the M67 models, Fe has diffused (downward) by 0.04 dex at 6000 K but by over 0.08 dex near 6100 K 
(see their Fig. 2). Just past the turnoff, the abundance-$T_{eff}$ relation loops even as dredge-up begins because the SCZ is shallower as 
compared to the depth of a pre-turnoff dwarf of the same $T_{eff}$ (see also \citet{del90}). The abundance continues to rise with further 
evolution, and reaches the initial model value by the base of the RGB. (There is slightly less diffusion in the cooler turnoff models of NGC 188, 
as compared to the turnoff of M67, and it should be kept in mind that slow mixing, for example, as induced by rotation, can inhibit diffusive motions; 
see also \citet{sil00, del98}.)

It should be stressed that the super-metal-rich turnoff/dwarf stars of NGC 6253 are expected to be more massive than those in M67 (at a given $T_{eff}$),
which may lead to different diffusion model predictions.  Nonetheless, the above predictions are consistent with our observations of Fe in NGC 6253.
As discussed above, the cluster Fe abundances seem to decrease from the coolest (faintest) dwarfs near $V$ =16.5 up to the turnoff, by of order
0.1 dex or so, though the Fe-$V$ slope is significant at only the 2.7 $\sigma$ level (and could be an artifact of the reduction procedures). Undeterred
by the lackluster statistical support of this argument, we can attempt to examine the data even more closely.  It is tempting to conclude that most of 
the slope is indeed seen in Fig. 8 where predicted, namely from $V$ =16.5 up to the top of the purely vertical part of the main sequence at $V$ = 15.3. 
Above this, stars become slightly cooler (up to $V$ = 15.0), where there is somewhat of a turnoff gap (up to about $V$ = 14.6), and then we see the very
hottest $subgiants$.  The model predictions are complex in this region (see the bottom of the Fe-$T_{eff}$ curve in Fig. 2 of \citet{mic04}), 
with clear dredge-up in the cooler subgiant models.  This is not inconsistent with our Fe data from $V$ = 15.3 to 14.5, including the seemingly 
higher Fe abundances in our brightest two stars, although the error bars are also consistent with no real scatter. Finally, although our average [Fe/H] 
of $+0.46^{0.02}_{0.03}$ for the giants (see Sec. 4.3) is within the errors of our value of $+0.43 \pm 0.01$, for the dwarfs, a small difference between the two, as 
perhaps predicted by diffusion, is possible.  In short, our data are not inconsistent with diffusion, and are even slightly suggestive of its effects, 
but with only a weak statistical significance, at best.

It is interesting to point out that, although the metallicities reported in the various studies of NGC 6791 (see Sec. 1) essentially agree to
within the errors, the lowest reported spectroscopic value comes from the two turnoff stars of \citet{boe09}; all other values are higher, and are from
stars in more advanced stages of stellar evolution.

\subsection{Giant Abundances - Fe}
We approached the abundance determinations for the giant stars somewhat differently for a number of reasons. First, while our target selection among 
the MS/TO candidate stars was optimized by photometry to exclude non-members, the configuration involving brighter stars (giants, subgiants, potential blue 
stragglers) was designed to include any plausible candidate in the field. As a result, the fraction of non-members is higher. Perhaps more important, 
due to the high metallicity we expected spectra for the evolved stars to be at least as difficult to interpret as 
the MS/TO stars; tests with synthetic spectra confirmed that continuum placement would be challenging for the giant branch stars. Following 
\citet{car07}, we selected seven spectral regions surrounding iron lines for synthesis. Linelists were constructed using the 
online linelists of Kurucz focussed narrowly around the following iron lines: 6646, 6703, 6733, 6608/9, 6725/6, 6750/2 \AA. 
In addition, a 30 \AA\ region around the Li 6708 \AA\ line was synthesized and compared with observed spectra.  
For each member on the giant or subgiant branch at or below the level of the red giant branch clump, model atmospheres 
with the adopted effective temperature, log $g$ and microturbulent velocity were constructed for iron abundances 
ranging in 0.05 dex increments from 0.1 to 0.5, then compared by eye to the observed spectrum within the MOOG software suite. 
The iron abundance reported in Table 4 represents an average of the determinations from each separate wavelength 
region as determined by two separate measurers. The quoted error is the standard deviation. For the fourteen stars 
analyzed, the average of [Fe/H] values is +0.46 with a standard deviation of 0.03 (0.01 sem). We have been careful 
to maintain a distinction between averages of [Fe/H] values, as opposed to averaging non-logarithmic estimates of iron 
abundance; the logarithms of the average iron abundance yield an indistinguishable result for the representative average 
abundance of $+0.46^{+0.02}_{-0.03}$(sd).

Several probable non-members with CMD positions near the red giant branch clump also were analyzed in a similar way. 
If non-members, the assumptions leading to the adopted effective temperatures of the models would be incorrect. If, as is likely, the reddening correction 
is too large for foreground non-members, comparison to synthetic spectra with effective temperatures that are too high would lead to an over-estimate 
of the iron abundance. Abundances of [Fe/H] for these five stars range from 0.19 to 0.31, systematically lower than found among the
probable single-star members and consistent with an even lower abundance field population overcorrected for reddening. 

\subsection{Abundance Errors}
Systematic or random errors among any of several parametric decisions will lead to abundance errors.  For MS/TO stars, the dependencies of 
[Fe/H] on errors in surface gravity, effective temperature and microturbulent velocity were determined by exploring a grid of models used to analyze 
the measured equivalent width data, specifically the data for star 7292 in the vertical turnoff region typifying the MS/TO sample. In summary, 
the effects on [Fe/H] due to changes in $T_{eff}$ of $+100$ K, in log $g$ of 1.0, and in $V_t$ of -1.0 km/sec are found to be $0.08$, 
$0.09$, and $0.28$ respectively. A comparable exploration of model parameters for stars at the bright and faint end of the giant branch 
members supplies comparable sensitivies of $0.03$, $0.18$ and $0.46$ respectively, indicating even more modest sensitivity to temperature 
estimates but enhanced sensitivity to surface gravity and especially microturbulent velocity estimates.

For comparison, both \citet{ses07} and \citet{car07} present detailed explorations of the effects of parameter choices on their derived abundances. 
In summary, the effect on the [Fe/H] of a red giant branch star resulting from changes of $+100$ K in effective temperature and of $-1.0$ km/sec for 
microturbulent velocity are $+0.02$ and $+0.42$, respectively, similar to the sensitivities in our abundances for comparable stars. 
Examination of the effect of changing parameters on [Fe/H] for the MS/TO stars by \citet{ses07} may be compared with our results as follows: the 
effect on [Fe/H] of a $+100$ K change in effective temperature is +0.07 dex, while a change $-1.0$ km/sec in $V_t$ generates a shift of +0.30 dex in [Fe/H].

With the derived dependencies of [Fe/H] on each of our model parameters, it should be possible to estimate the systematic effect 
of different reddening or photometric zero-points; these dependencies may also be used to estimated the propagated effect of 
random photometric errors, if the forms of our color-temperature relation and formulation for $V_t$ may be assumed to be correct.  
The precision of $V$ and $(B-V)$ values for individual stars is 0.0031 mag and 0.0063 mag, respectively, for RG/BS stars and 0.0033 mag and 0.0083 mag, 
respectively, for MS/TO stars based upon the original photometry \citep{tatd}. As discussed in Sec. 2.3., comparison to the 
photometry of \citet{mon09} places a probable upper limit of 0.010 mag on the errors in $V$. A similar comparison in $B-V$ leads to a probable upper
limit to the errors of 0.012 mag. 

For MS/TO stars, the surface gravities may be determined with a precision of $\pm 0.05$ from isochrone comparisons, so the 
sensitivity to surface gravity uncertainties is negligible; for completeness, an error of 0.0033 mag in $V$ translates into a
random error of 0.001 in log $g$. A random error in $(B-V)$ of 0.0083 maps to a random error in
temperature of 27 K. Collectively, these affect $V_t$ at the level of 0.022 km/sec. Combined in quadrature, these random
errors imply a dispersion in [Fe/H] of $\pm$0.023. If we adopt the upper limits for the representative errors (0.010 mag and
0.012 mag in $V$ and $(B-V)$), the corresponding numbers for log $g$, $T_{eff}$, $V_t$, and [Fe/H] become 0.003, 39 K, 0.032 km/sec and 0.033 dex, respectively.

For the giants, $V_t$ depends only on log $g$, which again depends only on the $V$ magnitude. A random color error of 0.0063 mag folds
into an error in $T_{eff}$ of 12 K for the adopted color-$T_{eff}$ calibration. The effect on the log $g$ estimate due to the error in
$V$ is 0.002 in log $g$. In turn, this has an even smaller effect on $V_t$ of 0.00026 km/sec. Collectively, these produce a random error
in [Fe/H] below 0.01 dex. Raising the adopted uncertainties to 0.01 mag in $V$ and 0.012 mag in $(B-V)$ still leaves the random propagated
error in [Fe/H] at the 0.01 dex level.

\subsection{Comparison to Previous Work}
Abundance estimates from the present work may be compared directly to recent high dispersion spectroscopic 
studies by \citet{car07} and \citet{ses07}. The former study employs spectrum synthesis and a sample of giant stars to arrive 
at an [Fe/H] estimate for the cluster of +0.46. In contrast, \citet{ses07} explore a broader range of stellar type while using 
measured equivalent widths and abundances referenced to the sun, techniques similar to those employed in our study. Both of these 
studies analyze high signal-to-noise spectra ($\sim$100) of higher resolution ($\sim$45,000) than our dataset; both studies 
adopt a reddening value for NGC 6253 of E$(B-V)=0.23$, similar to the value employed here. Differences in adopted values for 
log $g$ were very minor and will not be addressed here.

A more detailed comparison to the results of \citet{car07} indicates some differences that belie the apparent similarity of their 
abundance result to ours. As noted earlier, from three of the probable member giants that are common to our survey, the effective temperatures 
adopted by \citet{car07} are $\sim 150$ K cooler than ours; their prescription for estimating microturbulent velocities leads to 
estimates $\sim 0.35$ km/sec lower than ours. Because the impacts of these two parameters on [Fe/H] shift in opposite senses, 
these choices partly offset each other in [Fe/H]. To facilitate a comparison of results, we will artificially adjust our abundance 
estimate in accordance with their parameter selection precepts; such an adjustment raises our average [Fe/H] estimate to [Fe/H] = +0.58. 

Comparison of our result to that of \citet{ses07} is complicated by various factors despite the apparent similarity of our approaches: 
abundances from equivalent widths within the MOOG framework referenced directly to solar spectra. None of the seven stars in 
the \citet{ses07} sample approaches the unevolved main sequence; of the two stars near the main sequence turnoff, one (2225) 
appears to be a non-member or SB candidate based upon its radial velocity. Additionally, as the authors themselves conclude, measuring 
equivalent widths in these highly metal-rich spectra is difficult enough for higher gravity stars, and extremely challenging for the giants. 
A potential complication is their inclusion of two apparent non-members among the sample of five evolved stars. Our radial velocities 
for 2542 and 3138 confirm the questionable membership of the former but indicate probable membership for the latter. Based on proper motions, 
both stars are probable members.

\citet{ses07} estimate effective temperature and microturbulent velocity from their
spectra. For the one probable member near the main sequence turnoff, their derived effective temperature is higher than our 
color-temperature estimation scheme would predict by $\sim 100$ K. Since the microturbulent velocity found for this star is lower than 
our estimation scheme would predict by $\sim 0.25$ km/sec, if we had adopted the temperature and microturbulent velocity derived by \citet{ses07} 
for this star, our derived abundance for this star would increase from [Fe/H] $= +0.45$ to $+0.59$. For the red giants, \citet{ses07} employ 
very similar effective temperatures to ours but, again, lower microturbulent velocities. If our abundances are rederived using the 
parameter estimates of \citet{ses07}, our abundances for these stars would be higher by 0.14 dex, or [Fe/H] = +0.60.

Several comments can be made about these comparisons. \citet{ses07} revise their own abundance estimates for two stars downward by 0.1 dex following 
a re-examination of several spectra using spectrum synthesis rather than equivalent widths. To counter what appears to be a suggestion that
synthesis produces lower abundances than equivalent width analysis, we offer the observation that both types of analysis were attempted for a
star at the base of the red giant branch, star 2562. Colors, surface gravity and microturbulent velocities were assigned to this star following the
procedures outlined for giant branch and clump stars. The S/N per pixel for the spectra is not high, resulting in a much larger than usual dispersion among the separate line indicators of iron abundance. Despite these issues, the line analysis of 14 lines produced [Fe/H] 
$= +0.48^{0.07}_{0.20}$ (sem), while 
spectrum synthesis led to [Fe/H] $=0.42 \pm 0.08$. On the basis of this one comparison star bridging the gravity range between the clump and 
turnoff stars, we can't provide any support for a methodological difference in abundance results for synthesis as compared to equivalent width measurements. 

Finally, we reiterate the important effect of averaging logarithmic abundances, as opposed to taking logarithms of abundance numbers. 
Particularly if the individual abundances include substantial scatter, averaging logarithmic values underestimates the more correct 
logarithm of the average numerical abundance; for our data the effect was 0.09 dex. The former practice is quite common; had 
we employed it, the average abundance for the large sample of MS/TO members would be 0.09 dex lower and more consistent with published results in 
the cluster once differences in parameter selections are accounted for. 

\section{Summary and Conclusions} 
High-dispersion spectra centered on the Li line of 89 potential members of the old open cluster NGC 6253 have been processed and analyzed, 
generating 65 probable radial-velocity members, including 47 dwarfs and 18 red giants. When coupled with recently published proper-motion
memberships, the member sample is reduced to 45 dwarfs and 18 red giants. Excluding potential binaries and spectra with inadequate 
S/N per pixel, abundances have been derived from line analysis for 38 stars at the turnoff and from spectrum synthesis for 15 red giants, 
with impressive agreement between the two samples for the key element, Fe. For the adopted reddening (E$(B-V)$ = 0.22) and effective temperature 
scale, the mean cluster metallicity lies at [Fe/H] = +0.45, confirming previous work from intermediate-band photometry of turnoff stars 
and high-dispersion analyses of a handful of mostly red giants that placed NGC 6253 among the most metal-rich systems known in the Galaxy,
if not the most metal-rich. 
The issues dogging any discussion of the metallicity of this object bear a strong similarity to those surrounding NGC 6791 - the exact 
value of the reddening, the appropriate color and effective temperature scale, and an embarrassing richness of lines that can confuse the 
interpretation of spectra more than enlighten. Despite these caveats, the choices made in setting the cluster parameters have been 
selected such that probable changes would raise our metallicity determination for NGC 6253. On an absolute scale, if the effective temperature 
and microturbulence scales had been set using the approaches found in previous high-dispersion work, our NGC 6253 metallicity would be 
increased by $\sim$0.1 dex.

We have presented results for three elements other than iron from the wavelength region around the lithium line. In spite of a very limited 
linelist, we have recovered essentially solar values for [A/Fe] for nickel and silicon. The implied abundance for [Ca/Fe] is surprising and 
at variance with the results of \citet{car07} and \citet{ses07}, both of whom recover an essentially scaled-solar abundance for calcium. 
Our Ca abundance results rest entirely on a single line at 6717 \AA. In contrast, both studies from 2007 employ a wider spectral region 
including 6 lines for the \citet{car07} study and 22 for the \citet{ses07} work. A direct comparison of measured equivalent widths for
MS/TO stars in our study to published equivalent widths for the 6717 \AA\ line indicates that our equivalent width measurements are consistent 
with those of \citet{ses07}, implying that our measurement procedures are not faulty but that the line itself might be seriously contaminated. 
To test this possibility, equivalent widths were synthesized using MOOG for a solar model and a model typical of the MS/TO region's parameters 
in NGC 6253. For the solar model, the 6717 \AA\ line includes a blended feature from Ti II that contributes less than 8$\%$ to the effective 
equivalent width of the line. For the NGC 6253 turnoff star, however, the Ti II line accounts for a 20$\%$ contamination of the line feature. 
A quick correction to the equivalent width to ``remove" the contamination from the Ti II line suggests an estimate for [Ca/H] that is nearly 0.3 dex 
lower, consistent with an essentially scaled-solar abundance for calcium relative to iron.

While additional elements, including O, will be the focus of an analysis of a different wavelength region \citep{mad10}, the pattern
emerging from all high-dispersion work done to date is that NGC 6253, like NGC 6791 \citep{car06, or06, car07, boe09}, exhibits approximately 
scaled-solar abundances for a majority of the elements. Note that for NGC 6791, the only element that shows statistically significant
deviations from a scaled-solar ratio in multiple analyses is C \citep{or06, car07}; other elements such as Ba and Ni may show deviations in one study, 
but not in another, or worse, deviations in the opposite sense. As noted by \citet{boe09}, the pattern of roughly scaled-solar abundance ratios is 
consistent within the errors with what is observed among very metal-rich field dwarfs found in the solar neighborhood with kinematics that imply 
a potential origin near the galactic bulge \citep{pom02}, as well as among probable bulge dwarfs studied
{\it in situ} through spectroscopy obtained during gravitational lensing events \citep{coh09}. If a link exists between the bulge and NGC 6253, it
will need to await additional observational evidence before becoming apparent. 

\acknowledgements
The authors wish to thank the staff at Cerro Tololo Inter-American Observatory for their invaluable assistance in obtaining the
observations that form the basis of this study. CTIO is operated by the Association of Universities for Research in Astronomy, under 
contract with the National Science Foundation. The discussion was greatly enhanced by the photometric and proper-motion data
made available by Dr. Montalto and his colleagues. Extensive use was made of the SIMBAD database, operating at CDS, Strasbourg, France and 
the WEBDA database maintained at the University of Vienna, Austria (http://www.univie.ac.at/webda). CPD gratefully acknowledges support from
the National Science Foundation under grant AST-0607567. BJAT and BAT are also grateful to the Astronomy Department 
at Indiana University for hospitality during our Fall 2008 stay and to the University of Kansas for sabbatical support. We all appreciate the
 thoughtful comments of the referee which improved the paper.

\clearpage
\figcaption[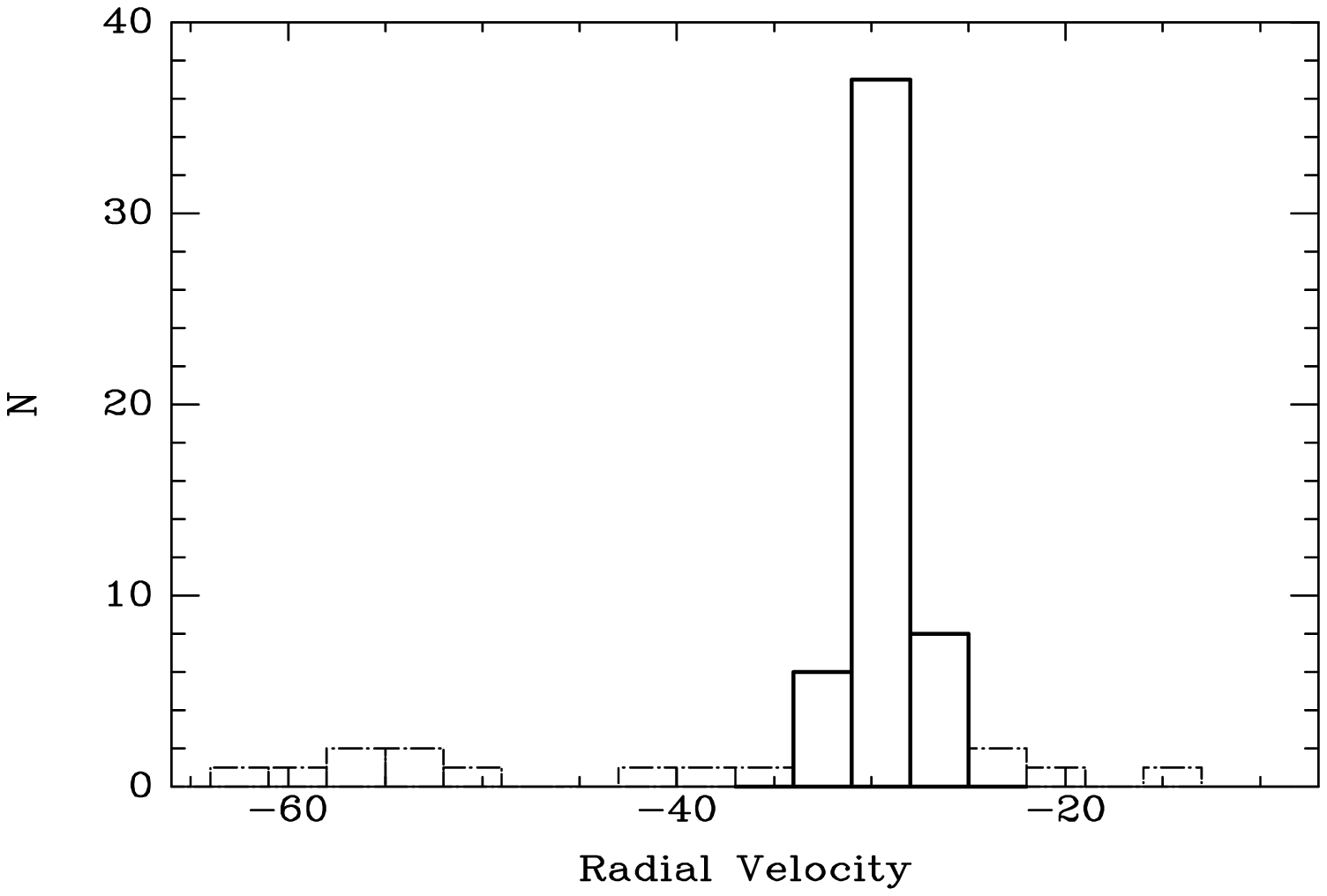]{Radial-velocity distribution for the stars in Table 1. The thick solid line identifies the velocity range selected 
for radial-velocity probable members. \label{f1}}

\figcaption[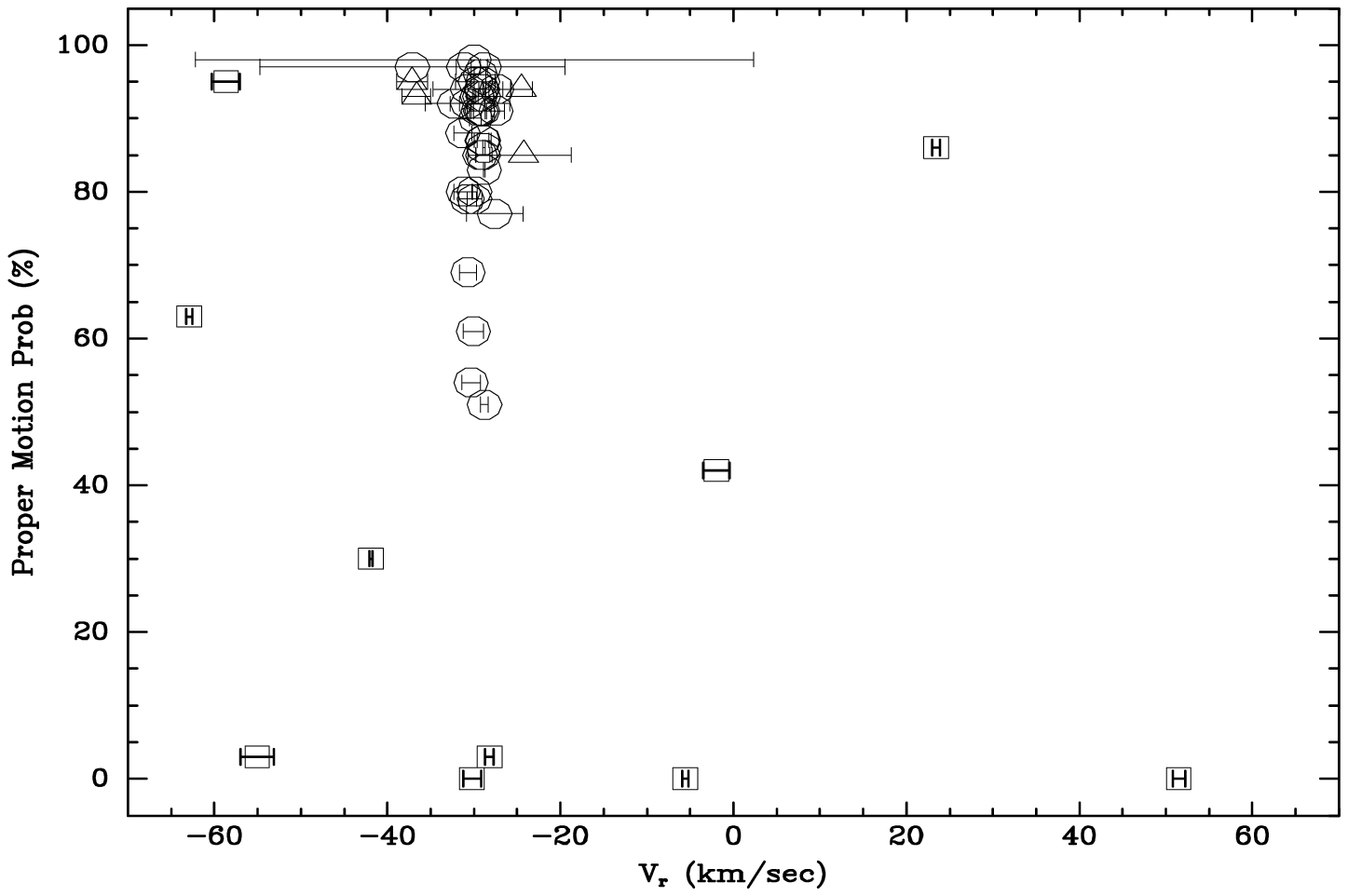]{Radial velocity as a function of proper-motion membership probability for the stars in Table 1. Circles
are stars tagged as likely single star members from the combined information. Squares are probable non-members or long period binaries, 
while triangles indicate stars with uncertain membership. \label{f2}}

\figcaption[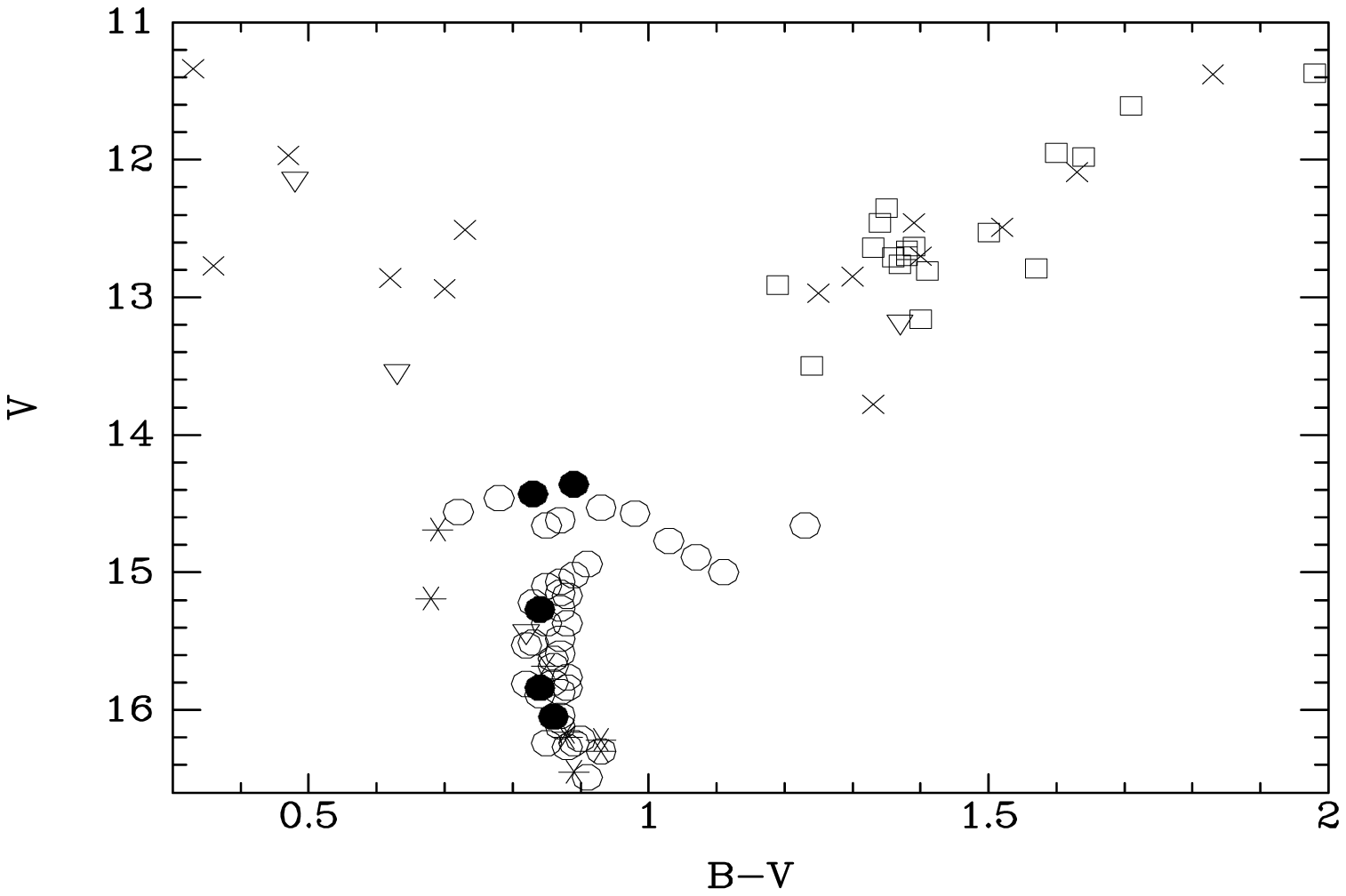]{CMD for the stars included in the current study - Table 1. Circles are MS/TO candidate stars; filled symbols are
probable binaries and asterisks are non-members. Squares are RG/BS candidate stars; open symbols are members and crosses are
probable non-members. Open triangles are stars with uncertain membership. \label{f3}}

\figcaption[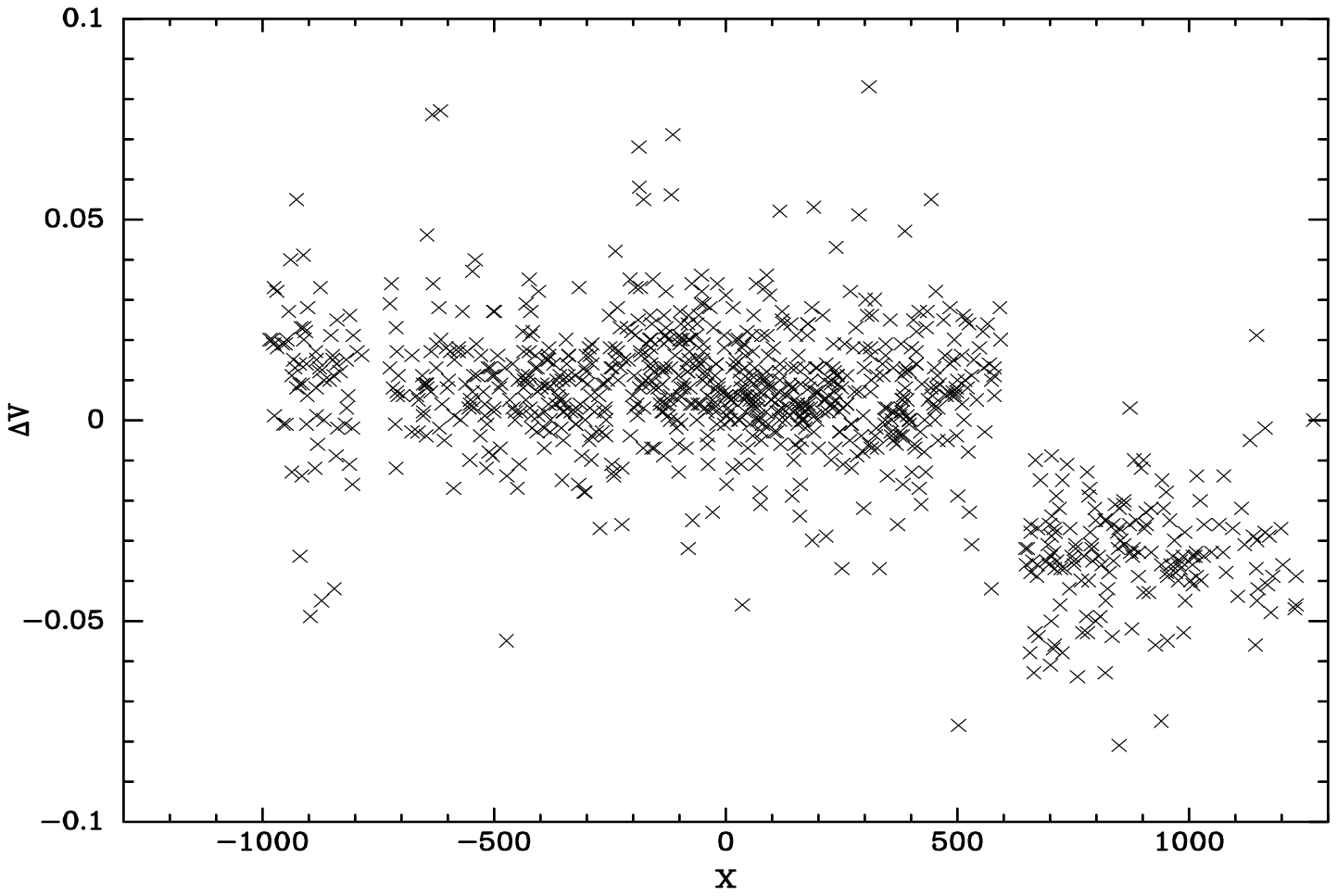]{(a) Residuals in $V$, in the sense (MON - TAD), for stars brighter than $V_{MON}$ = 16.5 as a function
of the X-coordinate position in WEBDA pixel coordinates. (b) Residuals in $V$, in the sense (MON - TAD), for stars brighter than $V_{MON}$ = 16.5 as a function of the Y-coordinate position in WEBDA pixel coordinates. \label{f4a}}

\figcaption[fig5.eps]{(a)The residuals in $B-V$, in the sense (MON - REF), as a function of $(B-V)_{MON}$ for the photometry of 
\citet{pia98} (crosses) and \citet{sag01} (open circles). (b) Same as Fig. 5a for the data of \citet{bra97}. \label{f5b}} 

\figcaption[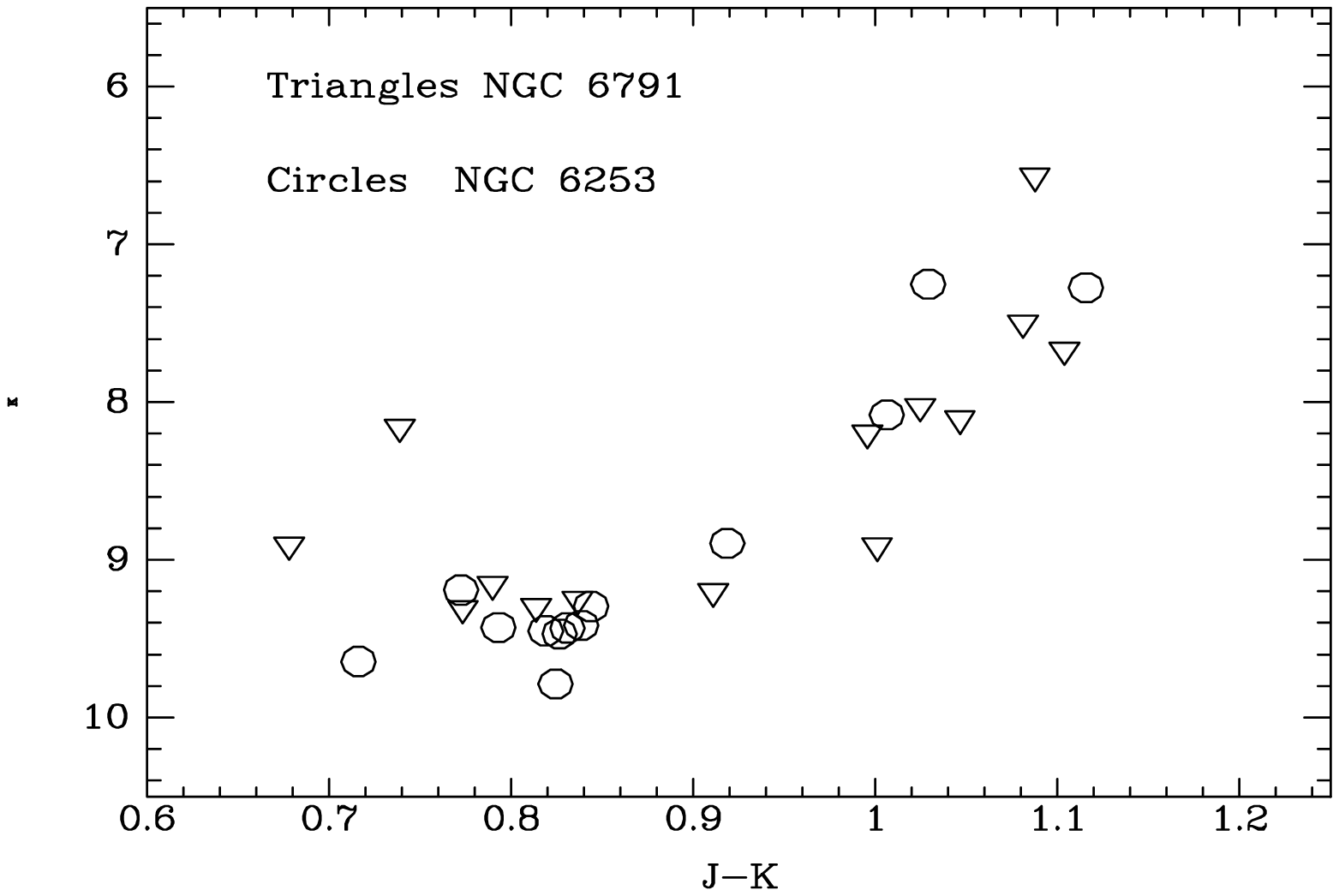]{The $JK$ CMD for stars in NGC 6791 (triangles) and NGC 6253 (circles). Giants in NGC 6791
have been shifted to the reddening and distance of NGC 6253 by applying offsets of $\Delta$$(J-K)$ = +0.06 mag
and $\Delta$$M_K$ = -2.1 mag to the individual points. \label{f6}}

\figcaption[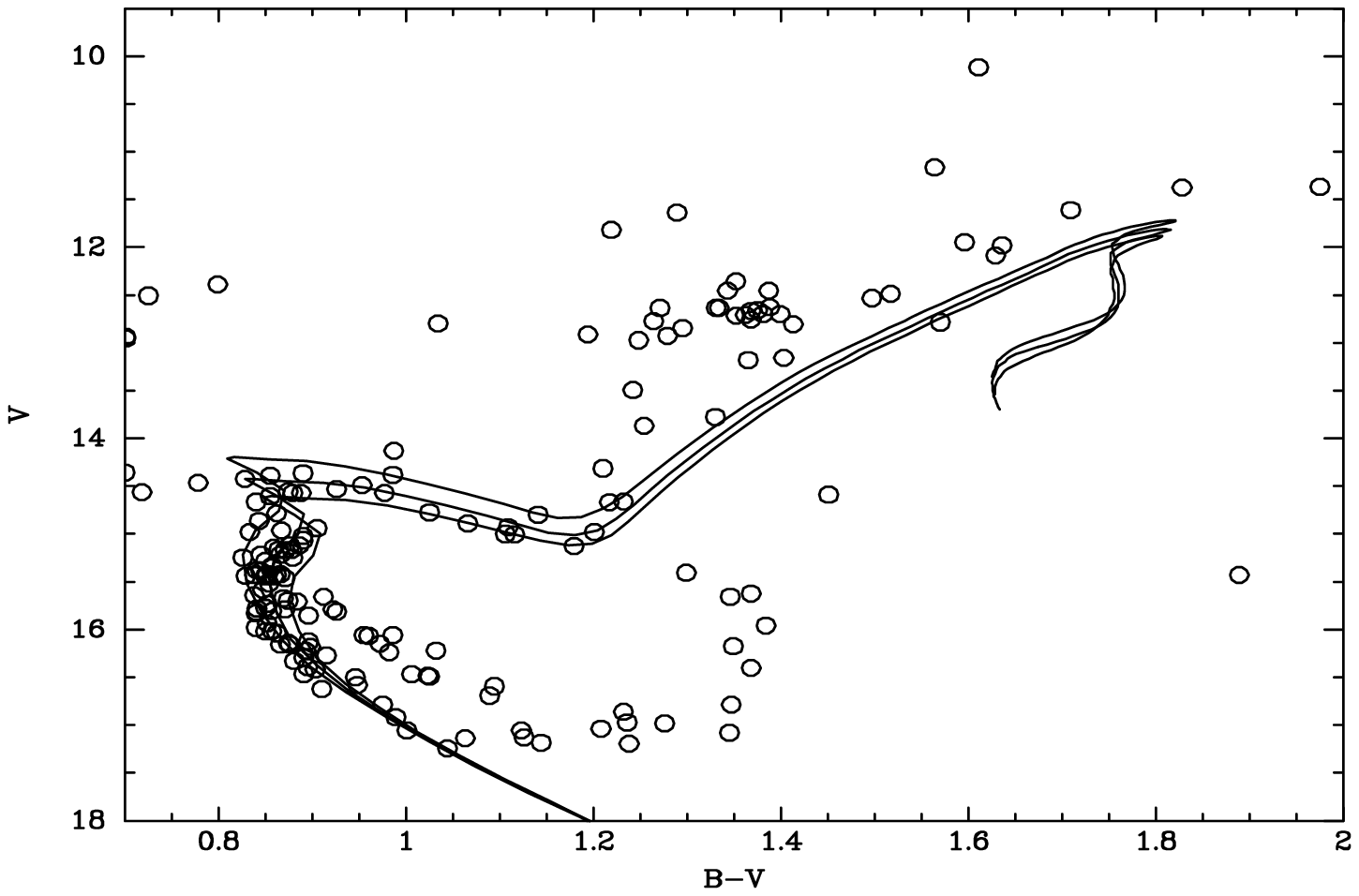]{Comparison of $BV$ photometry for NGC 6253, compiled by \citet{tatd}, to isochrones from
$Y^2$. Isochrones with a scaled-solar composition of have been adjusted for the apparent distance modulus of the cluster, $\mu = 11.9$ 
and the adopted reddening E$(B-V)=0.22$. Isochrones with ages 2.6, 3.0 and 3.4 Gyr are shown. \label{f7}}

\figcaption[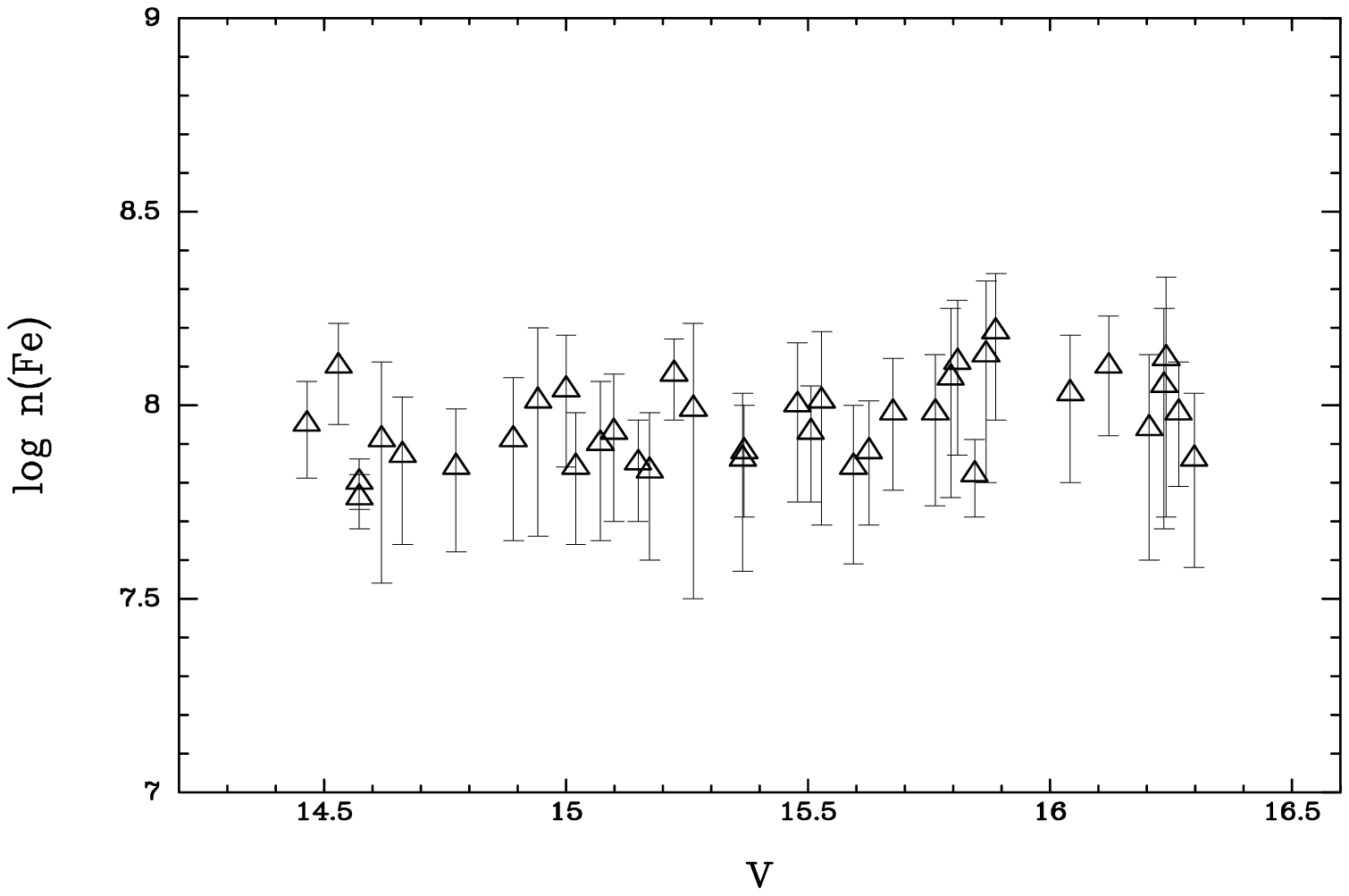]{Average Fe abundance for each MS/TO star as a function of the $V$ magnitude. Errors bars are the 
standard deviations in Fe for a given star. \label{f8}}

\figcaption[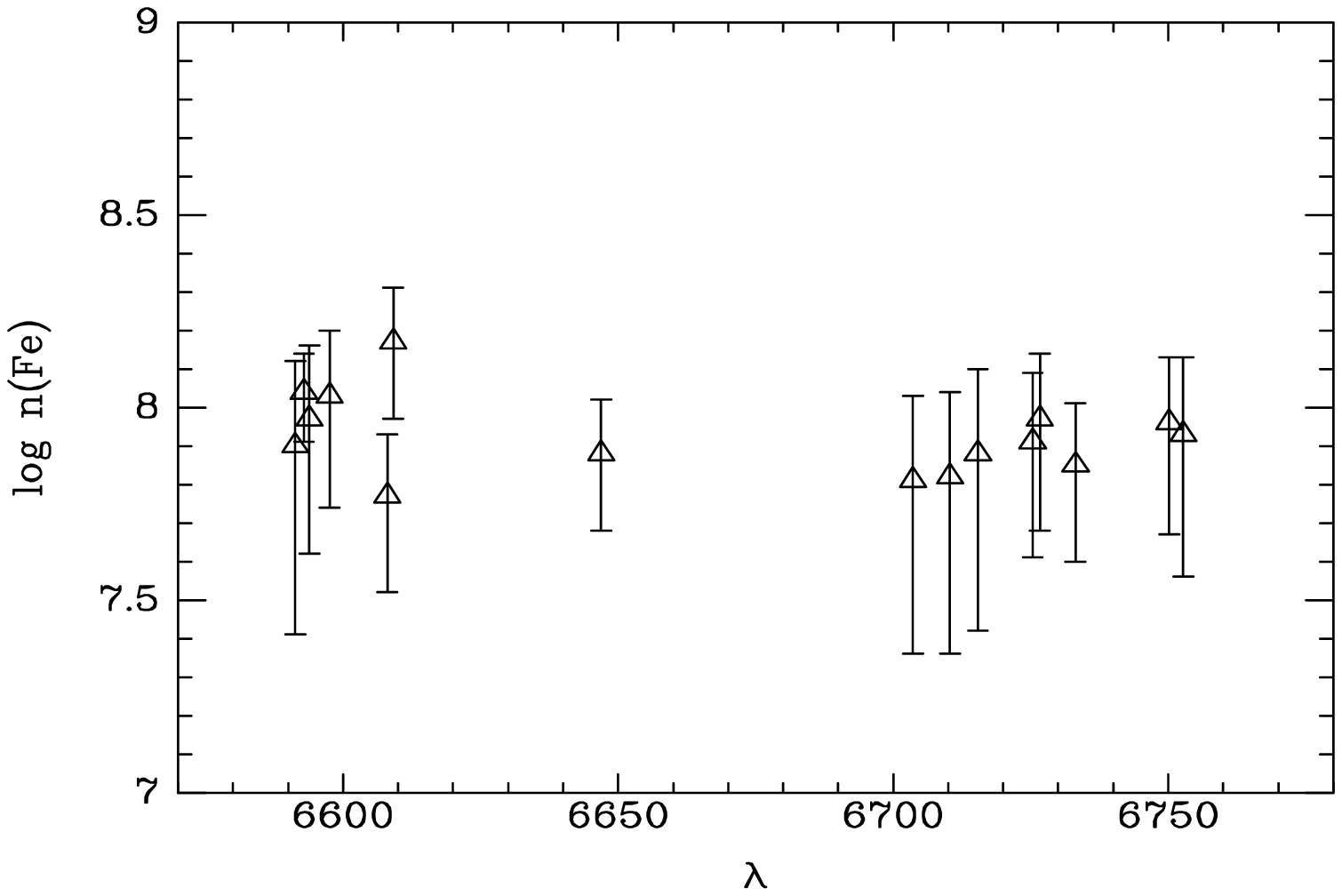]{Average Fe abundance among the MS/TO stars as a function of the wavelength of the line used to
determine the abundance. Error bars are the standard deviation in Fe among the stars for a given line. \label{f9}}

\newpage
\plotone{fig1.eps}
\plotone{fig2.eps}
\plotone{fig3.eps}
\clearpage
\plottwo{fig4a.eps}{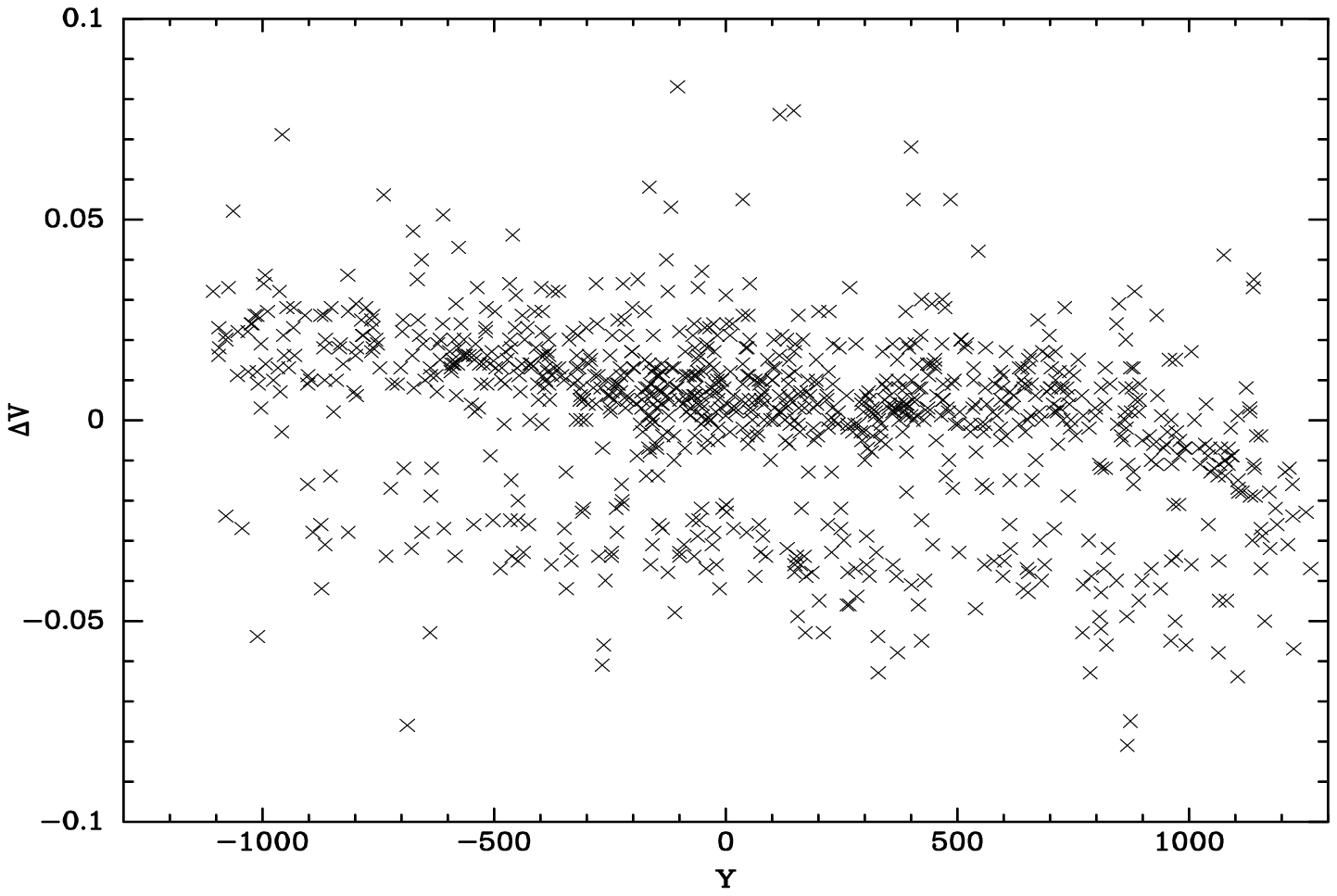}
\clearpage
\plottwo{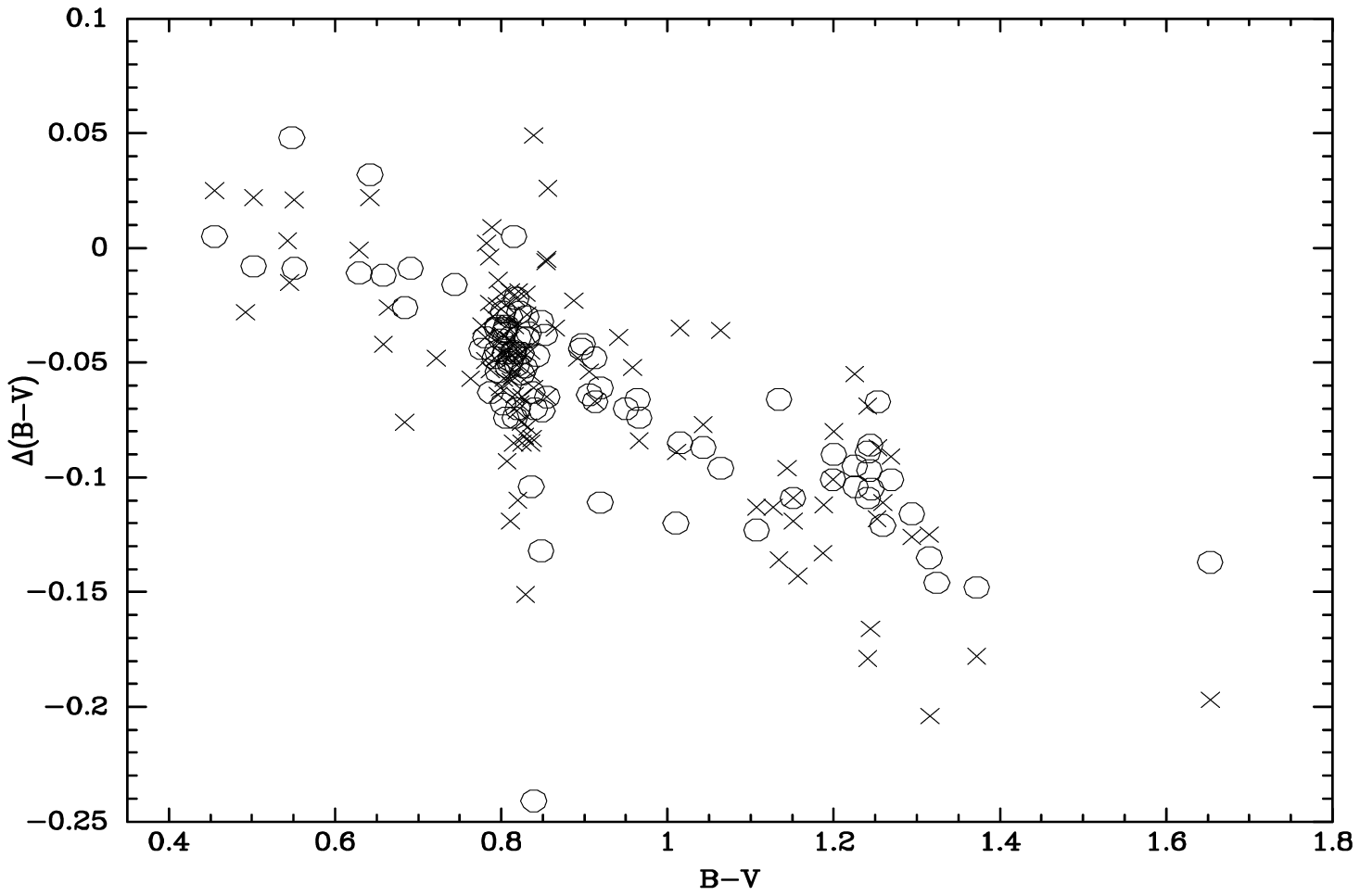}{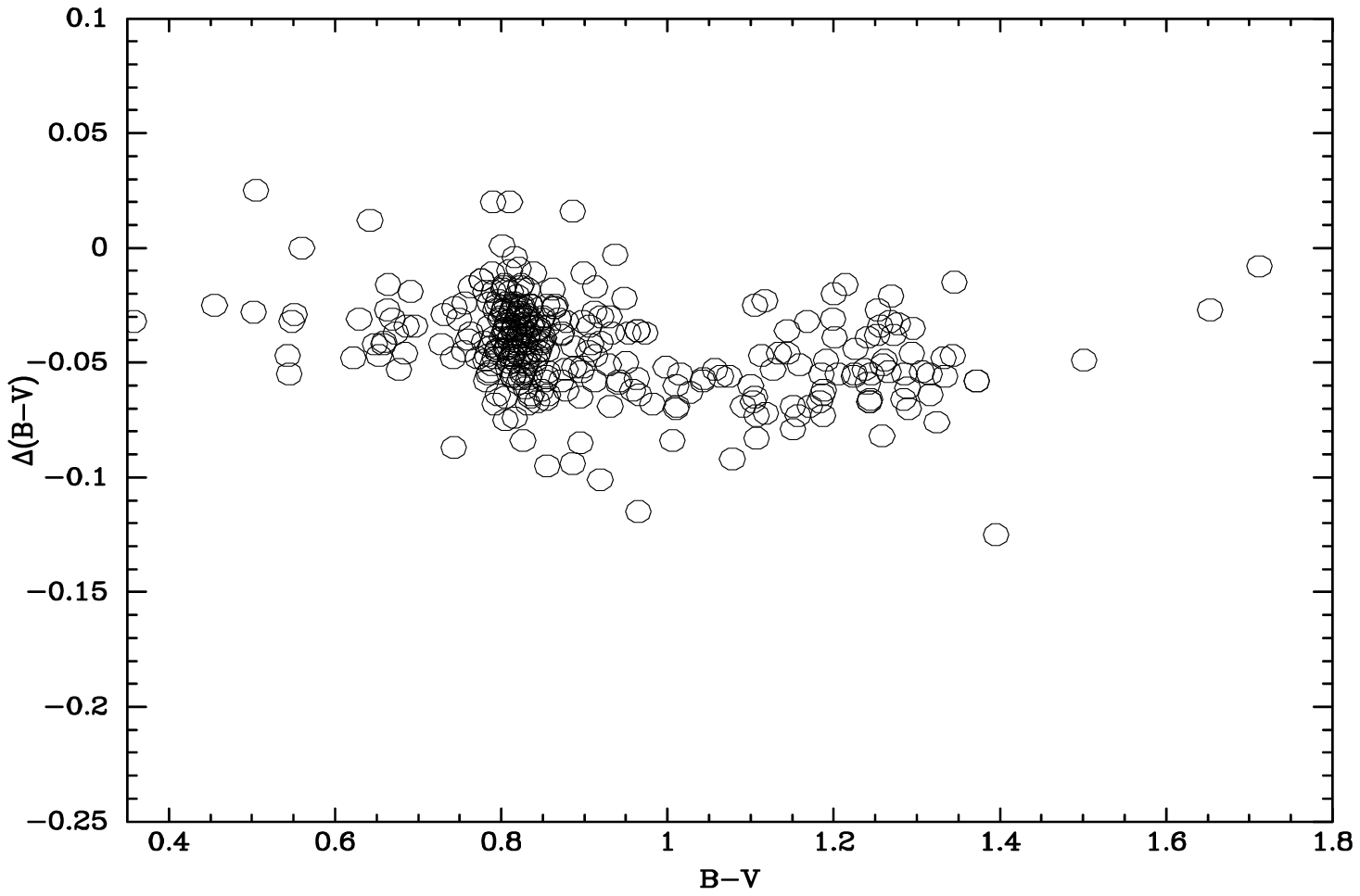}
\clearpage
\plotone{fig6.eps}
\plotone{fig7.eps}
\plotone{fig8.eps}
\plotone{fig9.eps}

%\documentclass[preprint]{aastex}
%\begin{document}
\begin{deluxetable}{crrcrrrrcrcc}
\pagestyle{empty}
\tablenum{1}
\tabletypesize\scriptsize
\tablecolumns{12}
\tablewidth{0pc}
\tablecaption{NGC 6253 Star Identifications and Radial Velocities}
\tablehead{
\colhead{WEBDA ID} &
\colhead{$V_{rad}$ } &
\colhead{$\sigma({V_r)}$ } &
\colhead{Num.} &
\colhead{$V$} &
\colhead{$V_{Mo}$} &
\colhead{$B-V$ } &
\colhead{$B-V_{Mo}$} &
\colhead{RVmem} &
\colhead{S/N } &
\colhead{P($\mu$)} &
\colhead{Memb} }
\startdata
 7019 & -13.78 & 3.75 & 1 & 11.34 & \nodata & 0.33 & \nodata & nm & 97 & \nodata & NM \cr
 1760 & -27.89 & 1.42 & 1 & 11.37 & \nodata & 1.98 & \nodata & m & 24 & \nodata & M \cr
 7020 & -49.43 & 1.25 & 1 & 11.38 & \nodata & 1.83 & \nodata & nm & 22 & \nodata & NM \cr
 1492 & -31.75 & 0.95 & 1 & 11.61 & \nodata & 1.71 & \nodata & m & 24 & \nodata & M \cr
 7023 & -29.61 & 0.90 & 1 & 11.95 & \nodata & 1.60 & \nodata & m & 25 & \nodata & M \cr
 7024 & -21.95 & 1.36 & 1 & 11.97 & \nodata & 0.47 & \nodata & nm & 127 & \nodata & NM \cr
 7025 & -29.58 & 0.87 & 1 & 11.98 & \nodata & 1.64 & \nodata & m & 23 & \nodata & M \cr
 7027 & 4.59 & 0.87 & 1 & 12.09 & \nodata & 1.63 & \nodata & nm & 27 & \nodata & NM \cr
 5201 & -24.25 & 5.52 & 1 & 12.14 & 12.148 & 0.48 & 0.492 & nm? & 82 & 85 & M? \cr
 3595 & -28.93 & 1.21 & 1 & 12.35 & 12.360 & 1.35 & 1.252 & m & 28 & 93 & M \cr
 7028 & 3.73 & 0.99 & 1 & 12.46 & \nodata & 1.39 & \nodata & nm & 42 & \nodata & NM \cr
 7029 & -28.52 & 1.15 & 1 & 12.46 & \nodata & 1.34 & \nodata & m & 27 & \nodata & M \cr
 7030 & -57.68 & 0.71 & 1 & 12.49 & 12.502 & 1.52 & 1.440 & nm & 45 & \nodata & NM \cr
 7031 & -54.29 & 1.90 & 1 & 12.51 & 12.390 & 0.73 & 0.872 & nm & 48 & \nodata & NM \cr
 2541 & -29.93 & 0.84 & 1 & 12.53 & 12.553 & 1.50 & 1.372 & m & 28 & 79 & M \cr
         &        &      &   &        &       &      &       &   &    &    &   \cr
 2885 & -29.01 & 1.01 & 1 & 12.63 & 12.629 & 1.39 & 1.289 & m & 28 & 91 & M \cr
 7032 & -31.22 & 1.06 & 1 & 12.64 & 12.638 & 1.33 & 1.255 & m & 27 & 80 & M \cr
 7033 & -28.83 & 1.10 & 1 & 12.66 & \nodata & 1.38 & \nodata & m & 27 & \nodata & M \cr
 7034 & -28.83 & 1.06 & 1 & 12.70 & 12.642 & 1.38 & 1.300 & m & 26 & \nodata & M \cr
 7035 & 5.28 & 1.17 & 1 & 12.70 & 12.656 & 1.40 & 1.320 & nm & 33 & \nodata & NM \cr
 4510 & -27.48 & 1.05 & 1 & 12.71 & 12.713 & 1.36 & 1.268 & m & 32 & 91 & M \cr
 7037 & -30.71 & 1.01 & 1 & 12.76 & 12.755 & 1.37 & 1.290 & m & 30 & 69 & M \cr
 7038 & 40.95 & 9.03 & 1 & 12.77 & 12.717 & 0.36 & 0.412 & nm & 49 & \nodata & NM \cr
 7040 & -27.50 & 0.89 & 1 & 12.79 & \nodata & 1.57 & \nodata & m & 24 & \nodata & M \cr
 7042 & -30.07 & 1.16 & 1 & 12.81 & 12.811 & 1.41 & 1.310 & m & 27 & 61 & M \cr
 7043 & -73.72 & 1.15 & 1 & 12.85 & 12.821 & 1.30 & 1.295 & nm & 33 & 27 & NM \cr
 7044 & -52.93 & 1.33 & 1 & 12.86 & 12.782 & 0.62 & 0.675 & nm & 65 & \nodata & NM \cr
 7045 & -32.97 & 0.81 & 1 & 12.91 & 12.854 & 1.19 & 1.136 & m & 46 & \nodata & M \cr
 2027 & -1.98 & 1.55 & 1 & 12.94 & 12.931 & 0.70 & 0.691 & nm & 79 & 42 & NM \cr
 7048 & 0.37 & 0.78 & 1 & 12.97 & 12.930 & 1.25 & 1.200 & nm & 49 & \nodata & NM \cr
         &        &      &   &        &       &      &       &   &    &    &   \cr
 2269 & -28.96 & 1.37 & 1 & 13.16 & 13.162 & 1.40 & 1.294 & m & 28 & 95 & M \cr
 2542 & -24.52 & 1.27 & 1 & 13.18 & 13.175 & 1.37 & 1.269 & nm? & 29 & 94 & M? \cr
 3138 & -31.17 & 0.83 & 1 & 13.50 & 13.501 & 1.24 & 1.144 & m & 36 & 97 & M \cr
 2126 & -37.14 & 1.79 & 1 & 13.54 & 13.516 & 0.63 & 0.642 & nm & 50 & 95 & M?BS \cr
 2568 & -58.66 & 1.61 & 1 & 13.78 & 13.799 & 1.33 & 1.253 & nm & 39 & 95 & NM \cr
 3168 & -29.95 & 32.26 & 3 & 14.36 & 14.363 & 0.89 & 0.688 & varVr & 145 & 98 & M \cr
 2507 & -37.11 & 17.65 & 3 & 14.43 & 14.405 & 0.83 & 0.776 & varVr & 113 & 97 & M \cr
 1741 & -28.88 & 0.17 & 2 & 14.46 & \nodata & 0.78 & \nodata & m & 106 & \nodata & M \cr
 2561 & -30.29 & 0.54 & 4 & 14.53 & \nodata & 0.93 & \nodata & m & 66 & \nodata & M \cr
 2193 & -27.01 & 4.84 & 4 & 14.56 & \nodata & 0.72 & \nodata & m & 171 & \nodata & M \cr
 2864 & -32.24 & 0.53 & 4 & 14.57 & 14.579 & 0.98 & 0.906 & m & 99 & 92 & M \cr
 7156 & -28.13 & 0.21 & 4 & 14.62 & 14.570 & 0.87 & 0.869 & m & 136 & \nodata & M \cr
 2131 & -30.73 & 4.86 & 4 & 14.66 & 14.668 & 0.85 & 0.813 & m & 146 & 92 & M \cr
 2562 & -28.84 & 0.38 & 4 & 14.66 & 14.665 & 1.23 & 1.134 & m & 45 & 97 & M \cr
 3212 & -5.54 & 0.34 & 3 & 14.69 & 14.692 & 0.69 & 0.669 & nm & 145 & \nodata & NM \cr
\tablebreak
 1172 & -29.63 & 0.12 & 4 & 14.77 & 14.783 & 1.03 & 0.941 & m & 100 & 93 & M \cr
 1704 & -29.26 & 0.67 & 4 & 14.89 & 14.885 & 1.07 & 0.966 & m & 80 & 96 & M \cr
 1562 & -29.82 & 2.28 & 3 & 14.94 & 14.939 & 0.91 & 0.855 & m & 131 & 95 & M \cr
 1311 & -29.04 & 0.39 & 4 & 15.00 & 15.007 & 1.11 & 1.015 & m & 68 & 94 & M \cr
 1888 & -30.72 & 4.03 & 4 & 15.02 & 15.033 & 0.89 & 0.840 & m & 124 & 94 & M \cr
 7255 & -29.31 & 0.62 & 4 & 15.07 & 15.085 & 0.87 & 0.836 & m & 118 & 93 & M \cr
 7267 & -30.35 & 2.89 & 4 & 15.10 & \nodata & 0.85 & \nodata & m & 128 & \nodata & M \cr
 4735 & -27.36 & 1.63 & 4 & 15.15 & 15.178 & 0.87 & 0.829 & m & 121 & 94 & M \cr
 7284 & -27.85 & 0.38 & 4 & 15.17 & \nodata & 0.88 & \nodata & m & 131 & \nodata & M \cr
 3248 & -55.08 & 1.94 & 3 & 15.19 & 15.187 & 0.68 & 0.664 & nm & 132 & 3 & NM \cr
 7292 & -29.28 & 1.41 & 4 & 15.22 & 15.231 & 0.83 & 0.826 & m & 117 & 85 & M \cr
 7300 & -29.19 & 0.49 & 2 & 15.26 & 15.232 & 0.87 & 0.854 & m & 84 & \nodata & M \cr
 7303 & -18.38 & 16.42 & 3 & 15.27 & 15.235 & 0.84 & 0.850 & varVr & 101 & \nodata & M \cr
 7329 & -29.16 & 0.27 & 4 & 15.37 & 15.329 & 0.88 & 0.903 & m & 82 & \nodata & M \cr
 596 & -29.13 & 0.87 & 3 & 15.37 & 13.379 & 0.85 & 0.793 & m & 110 & 94 & M \cr
         &        &      &   &        &       &      &       &   &    &    &   \cr
 7348 & -36.63 & 1.65 & 3 & 15.43 & 15.423 & 0.82 & 0.791 & nm & 127 & 93 & M? \cr
 7370 & -28.95 & 0.81 & 4 & 15.48 & 15.476 & 0.87 & 0.841 & m & 91 & 85 & M \cr
 7377 & -28.70 & 0.88 & 4 & 15.51 & 15.532 & 0.83 & 0.788 & m & 93 & \nodata & M \cr
 7390 & -30.90 & 0.96 & 4 & 15.53 & 15.495 & 0.82 & 0.834 & m & 93 & \nodata & M \cr
 3141 & -28.81 & 0.12 & 4 & 15.59 & 15.603 & 0.87 & 0.827 & m & 104 & 83 & M \cr
 7427 & -30.51 & 0.60 & 4 & 15.63 & 15.592 & 0.86 & 0.858 & m & 91 & \nodata & M \cr
 3643 & -30.77 & 1.03 & 4 & 15.68 & 15.679 & 0.86 & 0.830 & m & 108 & 79 & M \cr
 7448 & -62.90 & 0.37 & 3 & 15.68 & 15.672 & 0.85 & 0.847 & nm & 94 & 63 & NM \cr
 7470 & -29.73 & 3.24 & 4 & 15.76 & 15.738 & 0.88 & 0.873 & m & 93 & \nodata & M \cr
 4293 & -29.21 & 1.54 & 4 & 15.80 & 15.797 & 0.86 & 0.831 & m & 78 & 91 & M \cr
 7486 & -28.85 & 0.13 & 3 & 15.81 & 15.829 & 0.82 & 0.788 & m & 89 & 86 & M \cr
 7495 & -51.96 & 19.51 & 3 & 15.84 & 15.814 & 0.84 & 0.825 & varVr & 101 & \nodata & M \cr
 1474 & -29.51 & 0.82 & 4 & 15.84 & 15.857 & 0.88 & 0.825 & m & 114 & 91 & M \cr
 7505 & -31.27 & 1.01 & 4 & 15.87 & 15.893 & 0.87 & 0.819 & m & 65 & 88 & M \cr
 7511 & -28.93 & 0.65 & 4 & 15.89 & 15.906 & 0.84 & 0.810 & m & 68 & 87 & M \cr
         &        &      &   &        &       &      &       &   &    &    &   \cr
 7584 & -29.81 & 0.66 & 4 & 16.04 & 16.043 & 0.87 & 0.834 & m & 79 & 90 & M \cr
 7592 & -27.38 & 10.03 & 3 & 16.05 & 16.040 & 0.86 & 0.856 & varVr & 72 & \nodata & M \cr
 7627 & -29.97 & 1.33 & 4 & 16.12 & 16.082 & 0.87 & 0.870 & m & 71 & \nodata & M \cr
 7638 & -28.19 & 0.48 & 3 & 16.16 & 16.165 & 0.88 & 0.829 & m & 77 & 3 & NM \cr
 7662 & -41.90 & 0.21 & 2 & 16.20 & 16.201 & 0.88 & 0.855 & nm & 75 & 30 & NM \cr
 7664 & -29.01 & 0.99 & 4 & 16.21 & 16.221 & 0.90 & 0.854 & m & 77 & 87 & M \cr
 7672 & 23.42 & 0.48 & 3 & 16.22 & 16.298 & 0.93 & 0.825 & nm & 74 & 86 & NM \cr
 1691 & -28.77 & 0.44 & 4 & 16.24 & 16.249 & 0.89 & 0.849 & m & 72 & 51 & M \cr
 7676 & -30.35 & 1.07 & 4 & 16.24 & 16.241 & 0.85 & 0.840 & m & 45 & 54 & M \cr
 7694 & -27.60 & 3.30 & 4 & 16.27 & 16.279 & 0.88 & 0.855 & m & 83 & 77 & M \cr
 7718 & -25.39 & 0.79 & 4 & 16.30 & 16.322 & 0.93 & 0.841 & m & 70 & \nodata & M \cr
 7720 & 51.48 & 0.72 & 3 & 16.30 & 16.314 & 0.93 & 0.882 & nm & 76 & \nodata & NM \cr
 7806 & -30.23 & 1.03 & 4 & 16.45 & 16.470 & 0.89 & 0.852 & m & 57 & \nodata & NM \cr
 7822 & -29.94 & 0.33 & 3 & 16.49 & 16.496 & 0.91 & 0.865 & m & 34 & 80 & M \cr
\enddata
\tablecomments{S/N refers to the ratio of signal to noise per pixel; $m$ refers to probable member single stars; $nm$ may include both probable non-members and undetected binaries.} 
\end{deluxetable}
%\end{document}

\newpage
%\documentclass[preprint]{aastex}
%\begin{document}
\begin{deluxetable}{crrrrrrrrrrrrrrrrrrrr}
\pagestyle{empty}
\tablenum{2}
\tabletypesize\scriptsize
\setlength{\tabcolsep}{0.04in}
\rotate
\tablecolumns{21}
\tablewidth{0pc}
\tablecaption{Measured Equivalent Widths for MS/TO Stars}
\tablehead{
\colhead{WEBDA ID} & \multicolumn{15}{c}{Fe lines ($\lambda$ in \AA)} &
\multicolumn{3}{c}{Nickel lines} & \colhead{Ca line} & \colhead{Si line} \cr
 \cline{17-19} } 
\startdata
Wavelength & 6591.3 &  6592.9 & 6593.9 & 6597.6 & 6608.0 & 6609.1 & 6646.9  & 6703.6 & 6710.3 & 6715.4 & 6725.4  & 6726.7 & 6733.1 & 6750.2 & 6572.7  & 6643.6  & 6767.8 & 6772.3 & 6717.7  & 6721.8  \cr 
Exc. Pot & 4.59 & 2.73 & 2.43 & 4.79 & 2.28 & 2.56 & 2.61 & 2.76 & 1.48 & 4.61 & 4.10 & 4.61 & 4.64 & 2.42 & 4.64 & 1.68 & 1.83 & 3.66 & 2.71 & 5.86 \cr
log $gf$ & -1.81 & -1.30 & -2.23 & -0.90 & -3.93 & -2.57 & -3.87 & -2.98 & -4.71 & -1.38 & -2.16 & -1.03 & -1.33 & -2.65 & -1.18 & -1.92 & -2.02 & -0.86 & -0.26 & -1.07 \cr
\hline
1741 & 23.1 & 153.4 & 109.3 & 67.2 & 26.4 & 92.4 & 18.1 & 36.9 & \nodata & 37.0 & 19.3 & 60.9 & 41.6 & 93.1 & 43.3 & 123.3 & 89.6 & 62.8 & 148.2 & 65.7 \cr
2561 & 50.6 & \nodata & 129.0 & 72.5 & 57.9 & \nodata & 42.1 & 72.7 & \nodata & 71.1 & 44.1 & 81.6 & 50.2 & 104.4 & \nodata & 161.8 & 126.4 & 87.0 & \nodata & 80.6 \cr
2193 & \nodata & 127.1 & 83.5 & 47.1 & \nodata & 66.6 & \nodata & 33.8 & \nodata & 25.2 & 19.7 & 48.6 & 28.6 & 66.0 & 42.2 & 94.1 & 81.3 & 50.6 & 123.2 & 61.5  \cr
2864 & 30.9 & \nodata & 123.5 & 68.7 & 40.3 & \nodata & 28.1 & 58.2 & 42.9 & 54.0 & 36.6 & 71.4 & 46.2 & 96.4 & 63.0 & 134.5 & 109.1 & 73.3 & 166.7 & 73.4  \cr
7156 & 17.0 & 173.2 & 102.8 & 62.6 & 33.1 & 114.6 & 16.6 & 53.6 & 28.3 & 51.1 & 24.5 & 66.2 & 40.8 & 85.3 & 54.9 & 125.8 & 102.3 & 71.8 & 159.5 & 73.0 \cr
2131 & 21.3 & 157.4 & 110.9 & 62.3 & \nodata & 103.0 & \nodata & 50.5 & 26.7 & 35.1 & 22.9 & 53.1 & 44.2 & 89.8 & 51.0 & 116.7 & 106.8 & 67.1 & 153.8 & 76.2 \cr
1172 & \nodata & \nodata & 125.9 & 69.6 & 36.2 & 117.6 & 27.5 & 65.0 & 51.1 & 58.4 & 37.1 & 70.7 & 45.9 & 102.5 & 66.4 & 147.7 & 113.3 & 82.3 & 172.8 & 67.1 \cr
1704 & \nodata & 195.6 & 129.0 & 69.8 & 40.1 & 124.7 & 33.6 & 73.3 & 52.4 & 66.6 & 44.6 & 78.3 & 43.8 & 102.0 & 68.7 & 131.4 & 116.8 & 83.7 & 175.0 & 66.4 \cr
1562 & \nodata & \nodata & \nodata & \nodata & 18.9 & 112.6 & \nodata & 58.2 & 43.7 & 47.0 & \nodata & 80.2 & 58.9 & 90.9 & 55.3 & 122.9 & 113.4 & 73.5 & 169.5 & 77.4 \cr
1311 & 46.9 & \nodata & 147.2 & 78.5 & 46.8 & 127.4 & 41.0 & 75.9 & 70.8 & 70.4 & 46.0 & 78.8 & 53.8 & 117.5 & 84.2 & 151.2 & 127.2 & 78.1 & 189.7 & 69.8 \cr
1888 & 20.9 & 174.3 & 108.4 & 55.6 & 23.7 & 93.4 & \nodata & 41.2 & 23.1 & 40.6 & 33.5 & 65.4 & 37.5 & 85.6 & 57.2 & 118.2 & 100.4 & 65.8 & 156.0 & 64.1 \cr
7255 & 26.1 & 156.9 & 110.4 & 67.5 & 22.4 & 104.3 & \nodata & 43.3 & 23.1 & 48.0 & 32.9 & 66.5 & 43.3 & 80.3 & 55.4 & 121.0 & 106.1 & 70.3 & 160.0 & 77.9 \cr
7267 & 17.0 & 168.3 & 112.1 & 73.4 & 25.9 & 97.9 & \nodata & 41.2 & 26.7 & 40.8 & 31.8 & 66.8 & 31.8 & 89.3 & 47.4 & 116.5 & 96.1 & 67.2 & 146.4 & 75.1 \cr
4735 & 24.8 & 162.4 & 110.1 & 61.9 & 23.6 & 89.0 & 16.4 & 46.4 & 23.9 & 33.7 & 20.9 & 53.7 & 42.0 & 88.8 & 52.3 & 109.0 & 95.7 & 67.3 & 146.5 & 66.1 \cr
7284 & 17.7 & 155.1 & 92.0 & 67.4 & 19.3 & 95.9 & \nodata & 40.1 & 21.3 & 37.1 & 25.6 & 59.4 & 43.8 & 87.5 & 55.1 & 114.3 & 98.5 & 63.7 & 149.1 & 73.8 \cr
7292 & \nodata & 164.9 & 117.1 & 72.5 & \nodata & 80.5 & \nodata & 51.6 & 24.6 & 47.9 & 38.0 & 67.1 & \nodata & 86.0 & 56.4 & 121.6 & 92.2 & 69.5 & 173.0 & 81.9 \cr
7300 & 20.6 & 152.1 & 90.6 & 71.5 & 27.4 & 111.6 & \nodata & 47.7 & 28.2 & 47.2 & 29.0 & 53.6 & 41.1 & 96.7 & 62.8 & 125.1 & 121.0 & 82.7 & 161.3 & 82.0 \cr
7329 & 23.4 & 143.0 & 112.0 & 67.9 & 31.9 & 97.8 & \nodata & 46.0 & 17.3 & 32.0 & 24.1 & 52.2 & 33.2 & 83.9 & 58.5 & 112.0 & 98.0 & 69.0 & 145.9 & 70.2 \cr
596 & \nodata & 149.8 & 104.5 & 58.5 & 18.3 & 86.6 & \nodata & 43.0 & 29.1 & 27.8 & 25.3 & 56.5 & 43.8 & 92.7 & 50.4 & 113.8 & 97.4 & 70.8 & 145.2 & 65.9 \cr
7370 & \nodata & 168.4 & 120.8 & 52.7 & 22.0 & 96.4 & 19.8 & 45.7 & 17.0 & 51.8 & 28.5 & 71.5 & 58.9 & 94.0 & 59.3 & 121.6 & 83.3 & 67.1 & 148.7 & 73.0 \cr
7377 & \nodata & 161.3 & 93.1 & 61.4 & 15.5 & 80.0 & \nodata & 46.6 & \nodata & 50.8 & 25.2 & 68.6 & 43.5 & 70.5 & 46.5 & 103.4 & 96.9 & 58.9 & 143.2 & 66.5 \cr
7390 & \nodata & 142.2 & 121.6 & 55.5 & 24.3 & 97.8 & 20.7 & 41.1 & 16.0 & 54.0 & 27.9 & 66.3 & 39.4 & 76.0 & 50.9 & 103.3 & 91.1 & 83.7 & 142.0 & 59.2 \cr
3141 & 25.9 & 151.4 & 98.0 & 71.8 & 20.5 & 85.8 & \nodata & 37.1 & \nodata & 25.8 & 27.2 & 47.9 & 45.1 & 85.0 & 43.2 & 110.8 & 96.0 & 76.9 & 152.2 & 71.4 \cr
7247 & \nodata & 155.5 & 91.5 & 65.6 & 24.7 & 86.3 & \nodata & 38.8 & 21.1 & 37.1 & 33.2 & 66.2 & 26.4 & 80.6 & 46.3 & 116.1 & 95.2 & 57.9 & 139.1 & 73.1 \cr
3643 & 19.3 & 158.3 & 110.6 & 62.4 & 19.5 & 93.2 & 19.9 & 48.7 & 23.9 & 36.8 & 43.6 & 64.6 & 55.4 & 85.1 & 56.0 & 108.9 & 93.8 & 66.9 & 151.4 & 70.7 \cr
7470 & \nodata & 151.6 & 105.6 & 74.5 & 29.9 & \nodata & 15.3 & 47.6 & 19.9 & 45.8 & 45.2 & 79.5 & 52.1 & 96.1 & 52.6 & 120.8 & 113.3 & 70.5 & 160.9 & 74.7 \cr
4293 & 30.6 & 181.0 & 115.1 & 63.1 & 26.3 & 94.4 & 17.9 & 42.4 & 54.3 & 40.5 & 27.3 & 48.7 & 48.1 & 97.6 & 52.3 & 118.2 & 96.4 & 78.3 & 160.5 & 65.1  \cr
7486 & \nodata & 173.6 & 90.6 & 80.0 & 20.5 & 85.1 & \nodata & 55.3 & 23.1 & 36.7 & 36.6 & 80.9 & 52.4 & 84.3 & 46.4 & 113.0 & 112.6 & 57.6 & 147.8 & 89.7 \cr
1474 & \nodata & 152.1 & 93.8 & 58.2 & 21.0 & 83.3 & 16.9 & 48.3 & 27.5 & 33.1 & 20.9 & 62.1 & 41.2 & 82.5 & 52.0 & 119.7 & 101.6 & 64.7 & 151.8 & 69.1 \cr
7505 & \nodata & 156.3 & 87.4 & 62.5 & 21.9 & 97.8 & \nodata & 75.0 & 24.5 & 54.4 & 37.6 & 90.4 & 55.9 & 99.4 & 73.3 & 137.7 & 85.3 & 85.3 & 139.2 & 86.0 \cr
7511 & 48.5 & 166.2 & 131.6 & 77.5 & \nodata & \nodata & \nodata & 63.1 & 38.6 & 41.2 & 38.5 & 68.8 & 50.0 & 85.6 & 55.1 & 113.0 & 123.2 & 83.3 & 156.9 & 104.4 \cr
7584 & 19.3 & 174.6 & 95.6 & 81.0 & \nodata & 89.0 & 16.2 & 58.1 & 22.7 & 45.1 & 32.7 & 74.1 & 43.7 & 95.8 & 50.1 & 127.8 & 90.3 & 71.5 & 181.3 & 72.1 \cr
7627 & \nodata & 167.7 & 104.6 & 68.8 & 28.3 & 82.7 & 29.1 & 54.7 & 40.8 & 49.0 & 41.6 & 70.9 & 56.7 & 97.0 & 78.8 & 119.0 & 99.3 & 72.1 & 157.0 & 77.8 \cr
7664 & 28.6 & 175.2 & 97.2 & 78.9 & 26.0 & \nodata & \nodata & 48.5 & 36.8 & 68.0 & 23.0 & 65.7 & 38.0 & 76.7 & 52.7 & 124.3 & 94.7 & 84.4 & 147.7 & 58.1 \cr
1691 & \nodata & 163.0 & 118.3 & 91.9 & 32.7 & 86.8 & 16.2 & 41.4 & 16.9 & 52.1 & 54.3 & 77.9 & 50.6 & 91.8 & 55.5 & 131.4 & 102.1 & 60.5 & 175.2 & 85.7 \cr
7676 & \nodata & 133.6 & \nodata & 81.3 & 20.6 & 77.6 & 18.8 & 65.9 & 27.9 & 71.1 &  & 62.5 &  & 105.5 & 50.9 & 142.6 & 117.9 & 67.0 & 154.6 & 58.4 \cr
7694 & 21.7 & 175.8 & 109.4 & 72.9 & 40.0 & \nodata & 18.2 & 55.6 & 19.9 & 29.3 & 44.3 & 71.7 & 57.5 & 83.1 & 55.8 & 129.7 & 95.4 & 72.7 & 132.3 & 62.1 \cr
7718 & \nodata & 156.9 & 110.4 & 67.5 & 22.4 & 104.3 & \nodata & 43.3 & 23.1 & 48.0 & 32.9 & 66.5 & 43.3 & 80.3 & 55.4 & 121.0 & 106.1 & 70.3 & 160.0 & 77.9 \cr
\enddata
\tablecomments{Equivalent widths are expressed in milli{\AA}ngstroms (m\AA), excitation potentials in electron volts.}
\end{deluxetable}
%\end{document}

\newpage
%\documentclass[preprint]{aastex}
%\begin{document}
\begin{deluxetable}{ccccccccccccccc}
\pagestyle{empty}
\tabletypesize{\scriptsize}
\tablenum{3}
\tablecolumns{15}
\tablewidth{0pc}
\tablecaption{Abundances for TO/MS Stars in NGC 6253}
\tablehead{
\colhead{WEBDA ID} & \colhead{[Fe/H]} & 
\colhead{$s+$} & \colhead{$s-$} & 
\colhead{No. lines} & \colhead{Ni} & 
\colhead{$s+$} & \colhead{$s-$} & 
\colhead{Ca} & \colhead{$\sigma$} & 
\colhead{Si} & \colhead{$\sigma$} & 
\colhead{$T_{eff}$} & 
\colhead{log $g$ } & \colhead{$V_t$} }
\startdata
1741 & 0.43 & 0.03 & 0.04 & 14 & 0.42 & 0.12 & 0.16 & 0.69 & 0.06 & 0.34 & 0.06 & 6208 & 3.82 & 1.50  \cr
2561 & 0.58 & 0.04 & 0.04 & 11 & 0.80 & 0.10 & 0.13 & \nodata& \nodata & 0.50 & 0.09 & 5724 & 3.74 & 1.22  \cr
2193 & 0.24 & 0.02 & 0.02 & 11 & 0.21 & 0.01 & 0.01 & 0.42 & 0.05 & 0.34 & 0.04 & 6421 & 3.82 & 1.67  \cr
2864 & 0.28 & 0.02 & 0.02 & 13 & 0.37 & 0.07 & 0.08 & 0.60 & 0.04 & 0.41 & 0.06 & 5572 & 3.72 & 1.12  \cr
7156 & 0.39 & 0.06 & 0.07 & 15 & 0.39 & 0.06 & 0.07 & 0.69 & 0.04 & 0.40 & 0.04 & 5913 & 3.77 & 1.33  \cr
2131 & 0.35 & 0.05 & 0.05 & 13 & 0.41 & 0.04 & 0.05 & 0.66 & 0.04 & 0.45 & 0.05 & 5975 & 3.84 & 1.29  \cr
1172 & 0.32 & 0.05 & 0.05 & 13 & 0.51 & 0.08 & 0.09 & 0.58 & 0.03 & 0.34 & 0.06 & 5435 & 3.71 & 1.02  \cr
1704 & 0.39 & 0.05 & 0.06 & 14 & 0.40 & 0.04 & 0.05 & 0.53 & 0.04 & 0.36 & 0.08 & 5323 & 3.70 & 0.95  \cr
1562 & 0.49 & 0.08 & 0.09 & 9 & 0.50 & 0.04 & 0.04 & 0.73 & 0.04 & 0.47 & 0.06 & 5789 & 3.89 & 1.07  \cr
1311 & 0.52 & 0.04 & 0.05 & 14 & 0.54 & 0.05 & 0.06 & 0.57 & 0.04 & 0.45 & 0.09 & 5218 & 3.70 & 0.86  \cr
&   &         &      &     &       &      &        &       &   &     &     &      &   &     \cr
1888 & 0.32 & 0.04 & 0.05 & 14 & 0.36 & 0.05 & 0.05 & 0.63 & 0.04 & 0.28 & 0.05 & 5837 & 3.90 & 1.10  \cr
7255 & 0.38 & 0.05 & 0.06 & 14 & 0.48 & 0.04 & 0.04 & 0.70 & 0.04 & 0.48 & 0.05 & 5897 & 3.92 & 1.12  \cr
7267 & 0.41 & 0.05 & 0.05 & 14 & 0.39 & 0.05 & 0.06 & 0.58 & 0.04 & 0.45 & 0.05 & 5952 & 3.94 & 1.14  \cr
4735 & 0.33 & 0.03 & 0.04 & 15 & 0.34 & 0.01 & 0.01 & 0.56 & 0.05 & 0.32 & 0.05 & 5910 & 3.97 & 1.07  \cr
7284 & 0.31 & 0.05 & 0.05 & 14 & 0.38 & 0.05 & 0.05 & 0.58 & 0.04 & 0.43 & 0.05 & 5881 & 3.98 & 1.03  \cr
7292 & 0.56 & 0.03 & 0.03 & 11 & 0.52 & 0.09 & 0.12 & 0.88 & 0.03 & 0.55 & 0.06 & 6038 & 4.03 & 1.09  \cr
7300 & 0.47 & 0.07 & 0.09 & \nodata & 0.76 & 0.05 & 0.06 & 0.71 & 0.05 & 0.54 & 0.08 & 5910 & 4.02 & 1.00\cr
7329 & 0.34 & 0.06 & 0.06 & 14 & 0.44 & 0.01 & 0.01 & 0.52 & 0.06 & 0.39 & 0.07 & 5859 & 4.07 & 0.90  \cr
596  & 0.36 & 0.04 & 0.04 & 13 & 0.49 & 0.03 & 0.03 & 0.57 & 0.04 & 0.33 & 0.06 & 5971 & 4.07 & 0.99  \cr
7370 & 0.48 & 0.05 & 0.06 & 14 & 0.51 & 0.12 & 0.16 & 0.56 & 0.05 & 0.43 & 0.06 & 5894 & 4.11 & 0.87  \cr
&   &         &      &     &       &      &        &       &   &     &     &      &   &     \cr
7377 & 0.41 & 0.04 & 0.05 & 12 & 0.40 & 0.06 & 0.07 & 0.57 & 0.05 & 0.34 & 0.07 & 6021 & 4.12 & 0.96  \cr
7390 & 0.49 & 0.06 & 0.07 & 14 & 0.61 & 0.12 & 0.17 & 0.57 & 0.05 & 0.25 & 0.06 & 6055 & 4.13 & 0.97  \cr
3141 & 0.32 & 0.05 & 0.06 & 13 & 0.54 & 0.05 & 0.06 & 0.58 & 0.04 & 0.41 & 0.05 & 5888 & 4.16 & 0.80  \cr
7427 & 0.36 & 0.04 & 0.05 & 13 & 0.48 & 0.09 & 0.11 & 0.47 & 0.05 & 0.43 & 0.07 & 5920 & 4.17 & 0.81  \cr
3643 & 0.46 & 0.04 & 0.04 & 15 & 0.46 & 0.02 & 0.02 & 0.59 & 0.04 & 0.40 & 0.05 & 5926 & 4.18 & 0.81  \cr
7470 & 0.46 & 0.05 & 0.06 & 13 & 0.65 & 0.07 & 0.08 & 0.62 & 0.04 & 0.45 & 0.06 & 5865 & 4.21 & 0.80  \cr
4293 & 0.55 & 0.06 & 0.06 & 15 & 0.63 & 0.05 & 0.06 & 0.66 & 0.04 & 0.32 & 0.08 & 5932 & 4.21 & 0.80  \cr
7486 & 0.59 & 0.05 & 0.06 & 13 & 0.69 & 0.12 & 0.17 & 0.62 & 0.05 & 0.66 & 0.07 & 6061 & 4.22 & 0.86  \cr
1474 & 0.30 & 0.03 & 0.03 & 14 & 0.54 & 0.06 & 0.08 & 0.54 & 0.03 & 0.37 & 0.05 & 5874 & 4.23 & 0.80 \cr
7505 & 0.61 & 0.06 & 0.07 & 13 & 0.75 & 0.12 & 0.17 & 0.43 & 0.06 & 0.59 & 0.11 & 5894 & 4.24 & 0.80 \cr
&   &         &      &     &       &      &        &       &   &     &     &      &   &     \cr
7511 & 0.67 & 0.05 & 0.06 & 12 & 0.85 & 0.10 & 0.12 & 0.66 & 0.05 & 0.88 & 0.08 & 5995 & 4.24 & 0.80  \cr
7584 & 0.51 & 0.05 & 0.05 & 14 & 0.62 & 0.10 & 0.13 & 0.77 & 0.04 & 0.41 & 0.07 & 5910 & 4.27 & 0.80  \cr
7627 & 0.58 & 0.04 & 0.04 & 14 & 0.58 & 0.03 & 0.04 & 0.59 & 0.05 & 0.48 & 0.09 & 5913 & 4.28 & 0.80  \cr
7638 & 0.44 & 0.06 & 0.06 & 15 & 0.60 & 0.06 & 0.08 & 0.46 & 0.05 & 0.95 & 0.10 & 5872 & 4.29 & 0.80  \cr
7664 & 0.42 & 0.06 & 0.07 & 13 & 0.61 & 0.09 & 0.11 & 0.43 & 0.04 & 0.21 & 0.07 & 5811 & 4.30 & 0.80  \cr
1691 & 0.53 & 0.06 & 0.08 & 14 & 0.59 & 0.11 & 0.15 & 0.67 & 0.05 & 0.58 & 0.08 & 5849 & 4.31 & 0.80  \cr
7676 & 0.60 & 0.08 & 0.09 & 11 & 0.89 & 0.10 & 0.14 & 0.59 & 0.08 & 0.22 & 0.13 & 5971 & 4.31 & 0.80  \cr
7694 & 0.46 & 0.04 & 0.05 & 14 & 0.61 & 0.08 & 0.10 & 0.30 & 0.05 & 0.27 & 0.07 & 5856 & 4.31 & 0.80  \cr
7718 & 0.34 & 0.06 & 0.06 & 13 & 0.49 & 0.03 & 0.03 & 0.46 & 0.05 & 0.47 & 0.08 & 5725 & 4.32 & 0.80  \cr
7806 & 0.63 & 0.05 & 0.06 & 12 & 0.70 & 0.06 & 0.08 & 0.58 & 0.06 & 0.22 & 0.08 & 5849 & 4.35 & 0.80  \cr
\enddata
\end{deluxetable}
%\end{document}

\newpage
%\documentclass[preprint]{aastex}
%\begin{document}
\begin{deluxetable}{ccccccc}
\pagestyle{empty}
%\tabletypesize{\scriptsize}
\tablenum{4}
\tablecolumns{7}
\tablewidth{0pc}
\tablecaption{Iron Abundances from synthesis for RG Members}
\tablehead{
\colhead{WEBDA ID} & \colhead{[Fe/H]} & \colhead{$\sigma$} &
\colhead{N($\lambda$)} & 
\colhead{$T_{eff}$} & \colhead{log $g$ } & \colhead{$V_t$} }
\startdata
3595 & 0.46  &  0.04 &  7  &    4676 & 2.34 & 1.53  \cr
7029 & 0.46  &  0.04 &  6  &    4694 & 2.41 & 1.52  \cr
2541 & 0.49  &  0.02 &  7  &    4408 & 2.46 & 1.52  \cr
2885 & 0.46  &  0.06 &  7  &    4607 & 2.52 & 1.52  \cr
7032 & 0.45  &  0.03 &  6  &    4717 & 2.52 & 1.51  \cr
7033 & 0.46  &  0.04 &  7  &    4632 & 2.54 & 1.50  \cr
7034 & 0.46  &  0.04 &  6  &    4620 & 2.56 & 1.50  \cr
4510 & 0.48  &  0.04 &  7  &    4657 & 2.57 & 1.50  \cr
7037 & 0.40  &  0.07 &  7  &    4645 & 2.59 & 1.50  \cr
7040 & 0.46  &  0.06 &  4  &    4280 & 2.62 & 1.48  \cr
7042 & 0.49  &  0.02 &  7  &    4560 & 2.63 & 1.49  \cr
2269 & 0.46  &  0.04 &  6  &    4579 & 2.82 & 1.46  \cr
2542 & 0.43  &  0.06 &  6  &    4651 & 2.83 & 1.46  \cr
3138 & 0.41  &  0.04 &  6  &    4902 & 3.02 & 1.44  \cr
2562 & 0.45  &  0.03 &  6  &    4928 & 3.65 & 1.36  \cr
\enddata
\end{deluxetable}
%\end{document}

\newpage
%\documentclass[preprint]{aastex}
%\begin{document}
\begin{deluxetable}{cccccr}
\pagestyle{empty}
\tablenum{5}
%\tabletypesize\small
\tablecolumns{6}
\tablewidth{0pc}
\tablecaption{Abundances by line, MS/TO stars} 
\tablehead{
\colhead{$\lambda$} &
\colhead{No. stars} &
\colhead{[A/H] } &
\colhead{s.e.m.+ } &
\colhead{s.e.m.-} &
\colhead{Solar EW (m\AA) } }
\startdata
\cutinhead{Iron lines}
6591.3 & 20 & 0.36 & 0.07 & 0.08 & 16.4\cr
6592.9 & 33 & 0.52 & 0.02 & 0.02 & 146.9\cr
6593.9 & 36 & 0.44 & 0.04 & 0.04 & 98.4 \cr
6597.6 & 37 & 0.50 & 0.03 & 0.04 & 48.7 \cr
6608.0 & 33 & 0.26 & 0.03 & 0.04 & 21.2 \cr
6609.1 & 32 & 0.65 & 0.03 & 0.03 & 73.4 \cr
6646.9 & 19 & 0.36 & 0.04 & 0.04 & 13.0 \cr
6703.6 & 38 & 0.29 & 0.04 & 0.05 & 42.7 \cr
6710.3 & 33 & 0.30 & 0.05 & 0.05 & 22.3 \cr
6715.4 & 38 & 0.36 & 0.05 & 0.05 & 33.0 \cr
6725.4 & 36 & 0.38 & 0.04 & 0.04 & 20.8 \cr
6726.7 & 38 & 0.45 & 0.03 & 0.04 & 51.0 \cr
6733.1 & 36 & 0.33 & 0.03 & 0.03 & 33.8 \cr
6750.2 & 38 & 0.43 & 0.03 & 0.04 & 77.3 \cr
6752.7 & 37 & 0.40 & 0.03 & 0.04 & 41.7 \cr
\cutinhead{Calcium}
6717.7 & 37 & 0.61 & 0.02 & 0.02 & 127.0 \cr
\cutinhead{Silicon}
6721.8 & 38 & 0.43 & 0.03 & 0.03 & 48.8 \cr
\cutinhead{Nickel}
6643.6 & 38 & 0.60 & 0.03 & 0.03 & 102.0\cr
6767.8 & 38 & 0.50 & 0.04 & 0.04 & 88.3 \cr
6772.3 & 38 & 0.46 & 0.03 & 0.03 & 55.8 \cr
\enddata
\end{deluxetable}
%\end{document}

\end{document}